\documentclass[twocolumn,showpacs,aps,prl,superscriptaddress]{revtex4}
\usepackage[dvips]{graphicx}
\usepackage{dcolumn}
\usepackage{bm}
\usepackage{multirow}

\begin{document}

\title{Radiation from  a Josephson STAR-emitter}
\author{Richard A. Klemm}\email{klemm@physics.ucf.edu}
\affiliation{Department of Physics, University of Central Florida, Orlando, FL 32816, USA}
\author{Kazuo Kadowaki}\email{kadowaki@ims.tsukuba.ac.jp}
\affiliation{Graduate School of Pure \& Applied Sciences, University of Tsukuba, 1-1-1, Tennodai, Tsukuba, Ibaraki 305-8573, Japan}

\date{\today}

\begin{abstract}
We calculate the angular dependence of the radiation-zone output power and electric polarization of stimulated terahertz amplified radiation (STAR) emitted from a $dc$ voltage applied across  cylindrical and rectangular stacks of intrinsic Josephson junctions.  During coherent emission,  a spatially uniform $ac$ Josephson current density  in the stack acts as a surface electric current density antenna source, leading to an harmonic radiation frequency spectrum,  as in experiment, but absent in all cavity models of cylindrical mesas. We assume that spatial fluctuations of the $ac$ Josephson current cause its fundamental mode to lock onto the lowest finite energy cylindrical cavity mode,  causing it to resonate,  leading to a non-uniform magnetic surface current density radiation source, and a non-trivial combined fundamental frequency output power with linear polarization for general radiation directions.  This combination of the two coherent sources may be fully or partially coherent.  The coupling of the higher $ac$ Josephson harmonics to other cylindrical cavity modes is shown to be very weak, at best, and that emission is also linearly polarized.  For the more complicated rectangular mesas, the lowest energy modes are apparently not excited, but the non-uniform $ac$ Josephson current can  excite the  harmonic  sequence of modes with spatial variation across the rectangular widths, leading to combined radiation output both for the fundamental and the higher harmonics, which combination also may be either fully or partially coherent. We also present a model of the  superconducting substrate, in which it acts as a perfect magnetic conductor, greatly reducing the STAR-emitter power and modifying its angular dependence, especially parallel to the substrate. This model can be tested quantitatively by measurements of the angular dependence  of the emitted radiation, especially that of the second harmonic emitted from cylindrical mesas.  Quantitative measurements of the amount of coherence could also prove enlightening.  Based upon this substrate model, existing Bi$_2$Sr$_2$CaCu$_2$O$_{8+\delta}$ crystals atop  perfect electric conductors could have  STAR-emitter power in excess of 5 mW, acceptable for many device applications.
\end{abstract}

\maketitle

\section{I. Introduction}
The recent discovery of coherent THz radiation emitted from mesas of the high-temperature superconductor Bi$_2$Sr$_2$CaCu$_2$O$_{8+\delta}$ (BSCCO) has caused a great deal of excitement\cite{Ozyuzer,Kadowaki}.
In these experiments, rectangular mesas were fashioned by Ar ion milling of a single crystal of BSCCO, with a Au layer covering the mesa's top, to which an electrical lead was attached, and two additional electrical leads were attached to the remaining BSCCO crystal substrate.  Typical rectangular mesa dimensions were $60\times300\times1\> \mu$m.  By applying a static ($dc$) voltage  $V$ across the mesa, the $ac$ Josephson effect was generated in each of the $N\>\sim 10^3$ junctions involved in the mesa, and coherent $ac$ Josephson radiation  at THz frequencies was emitted. The $ac$ Josephson relation
 \begin{eqnarray}
 \omega_J&=&\frac{2eV}{\hbar N},\label{nuJ}
 \end{eqnarray}
 where $2e$ is the magnitude of the Cooper pair charge and $\hbar$ is Planck's constant divided by $2\pi$. Equation (\ref{nuJ}) relates the $ac$ Josephson frequency $\nu_J=\omega_J/(2\pi)$ to the applied $V/N$.  To avoid confusion, here we write $k$ and ${\bm x}$ for the wave vectors and positions outside the mesa, and $k'$ and ${\bm x}'$ for the wave vectors and positions inside the mesa.  Since the frequencies are the same inside and outside the mesa, they are left unprimed.
\subsection{A. Rectangular mesas}
Since BSCCO is a stack of Josephson junctions, with atomically thin superconducting layers separated by thicker dielectric layers, it is extremely anisotropic.
In analogy with standard antenna theory\cite{antenna,antenna2,Jackson}, to zeroth order, one might consider a BSCCO mesa to be a dielectric sandwiched between two metallic layers, forming a cavity that produces the electric field that generates the radiation\cite{BK1,BK2,Koshelev,LinHu,LinHu2,LinHu3,LinHu4,HuLin,HuLin2,Matsumoto,Matsumoto2,LHT,Tachiki,Tachiki2,Koyama}.  In this model, it was usually assumed that the $ac$ component of the electric field ${\bm E}({\bm x}',t)$ and the $ac$ Josephson supercurrent ${\bm J}({\bm x}',t)$ both had antisymmetric spatial configurations, with maxima along the length $\ell$ edges, and a node in the center of the mesas of width $w$, leading to a radiation fundamental wavelength $\lambda=2w$, and a  finite magnetic field ${\bm H}({\bm x}',t)$ along the radiating mesa edges, as noted in the following. The wave vectors of very thin rectangular cavity modes are
\begin{eqnarray}
k'_{mp}&=&\pi\Bigl[\Bigl(\frac{m}{w}\Bigr)^2+\Bigl(\frac{p}{\ell}\Bigr)^2\Bigr]^{1/2},\label{krectangle}
\label{omegamp}
 \end{eqnarray}
 where   $m$ and $p$  are integers, when there is no spatial variation of the electromagnetic fields in the $\hat{\bm z}$ direction normal to the layers. These modes are generated from the boundary conditions, which allow for half-integral multiples of wavelengths across  the width and the length of the cavity. Usually, such modes in a cavity that is open on the sides only arises when the $ac$ magnetic field ${\bm H}({\bm x}',t)$ vanishes at the edge of the cavity, forcing the normal derivative of ${\bm E}({\bm x}',t)$ to vanish there.  Other mechanisms for the formation of such half-integral modes have also been proposed, as discussed in the following.  Inside the mesa, electromagnetic waves propagate parallel to the layers according to the dispersion relation
 \begin{eqnarray}
 \omega_{mp}&=&\frac{ck'_{mp}}{n_r},\label{omegarectangle}
 \end{eqnarray}
 where $n_r\approx\sqrt{\epsilon}$ is the index of refraction, and $\epsilon\approx18$ for BSCCO  in the relevant frequency range\cite{Ozyuzer,Kadowaki}.
 By varying the $dc$ $V$ experimentally, radiation is found to occur after one of the harmonic $ac$ Josephson frequencies $n\nu_J$ locks onto a particular $(m_0p_0)$ rectangular cavity mode,
 \begin{eqnarray}
 n\omega_J&=&\omega_{m_0p_0}=\frac{ck'_{m_0p_0}}{n_r}.
 \end{eqnarray}
   In  experiments on rectangular samples, it appears that the fundamental $n=1$ $ac$ Josephson  frequency $\nu_J$ locks onto the rectangular cavity $(10)$ mode, so that $\nu_J=c/(2n_rw)$, also allowing for the  higher $ac$ Josephson harmonics  with $n>1$ to lock onto the higher cavity $(m0)$ modes, where $m=n$.

  One might wonder why the low energy excitations were not along the length of the rectangular mesa, rather than along the width, as observed experimentally.   Although the experiments on rectangular mesas provided evidence for radiation frequencies $\nu_{m0}\propto m/w$  up to four harmonics of $\nu_J$\cite{Ozyuzer,Kadowaki}, a lower set of modes with harmonic frequencies would be $\nu_{0p}\propto p/{\ell}$, especially for $w\ll{\ell}$. Very recent experiments suggested that standing waves along the lengths of rectangular mesas might in fact occur\cite{Kleiner}, although it is presently uncertain if these modes radiate.  A possible resolution of this apparent paradox was recently suggested to arise from the  magnetic field component normal to the rectangular length, which was predicted to have a half-wavelength spatial variation across the rectangular width\cite{Tachiki}.

 However, this cavity analogy neglects the $ac$ Josephson supercurrent ${\bm J}({\bm x}',t)$ across each of the junctions.  Therefore, the second zeroth order model is to treat the mesa as a conducting dipole antenna with an $ac$ current source, as commonly used in microwave relay stations\cite{antenna,antenna2,Matsumoto,Matsumoto2,Tachiki,KK}.  In this picture, the most important part of the $ac$ Josephson current is its  portion that is uniform across the mesa.  This portion leads to a non-vanishing ${\bm H}({\bm x}',t)$ within the mesa  arising from Amp{\`e}re's law.   The important question then, is how to treat the Amp{\`e}re boundary condition.  This has been done in very different ways, with inconsistent results.
\subsection{B. The Amp{\`e}re boundary condition}
Bulaevskii and Koshelev studied  long rectangular mesas, and set the magnetic induction ${\bm B}$ parallel to each edge, with ${\bm E}||\hat{\bm z}$, normal to the substrate\cite{BK1,BK2}.  They then took $|{\bm B}/{\bm E}|=\pm\zeta_{\omega}(k_z)$ at the sample surface, where $\zeta_{\omega}=1$ for $|k_z|<2\pi\nu_J/c=k_{\omega}$, and $\zeta_{\omega}$ was imaginary and strongly $k_z$-dependent for $k_z>k_{\omega}$.  The same sort of model for the boundary conditions was studied with $\zeta_{\omega}=Z(\omega)$, a complex constant, by Lin and Hu\cite{LinHu,LinHu2,LinHu3}, a frequency-independent constant by Lin, Hu, and Tachiki\cite{LHT}, and $\zeta_{\omega}=\gamma=0.1$ by Tachiki {\it et al.}\cite{Tachiki}.  Hu and Lin properly took ${\bm B}=0$ on the edge of cylindrical mesas, but did not properly take account of the $ac$ Josephson current in the Amp{\`e}re boundary condition\cite{HuLin,HuLin2}.  Matsumoto {\it et al.} treated the Amp{\'e}re boundary condition as providing constraints on ${\bm B}$ on opposite sides of the long rectangular mesas in their numerical studies\cite{Matsumoto,Matsumoto2}.   Koyama {\it et al.} treated the $ac$ Josephson current as uniform in the mesa\cite{Koyama}.  The general problem with this model for the boundary conditions is that  a  clear understanding of either $\gamma$ or $\zeta_{\omega}(k_z)$ was not provided.  In the radiation zone, one expects $\zeta_{\omega}=\gamma=\pm1$, but at the edge  of the sample, ${\bm B}$ is given by the Amp{\'e}re boundary condition, which depends upon the $ac$ Josephson current within the mesa.  Thus, the actual situation appears at first sight to be much more complicated than as suggested in those papers.

Here we study the simplest and least ambiguous way to treat this boundary condition. We use Love's equivalence principles\cite{antenna,antenna2,Schelkunoff}, which are variations on Huygen's principle, treating the mesa as both an electric conductor and as a magnetic conductor\cite{antenna,antenna2}.  That is, the electric field within  the mesa is treated as a surface magnetic current density source ${\bm M}_S({\bm x}',t)$ by  the electric conductor equivalence principle, and the magnetic field inside the mesa is treated as a surface electric current density source ${\bm J}_S({\bm x}',t)$, using the magnetic conductor equivalence principle\cite{antenna,antenna2}.  A sketch of a cylindrical mesa with a uniform ${\bm J}_S$ and a nonuniform ${\bm M}_S$  corresponding to the $(11)$ cylindrical cavity mode is shown in Fig. 1.  This combination of equivalence principles allows us to obtain analytic forms for the angle dependence and polarization of the radiation  emitted from cavities suspended in vacuum, while properly accounting for the Amp{\'e}re boundary condition.
\begin{figure}
\includegraphics[width=0.45\textwidth]{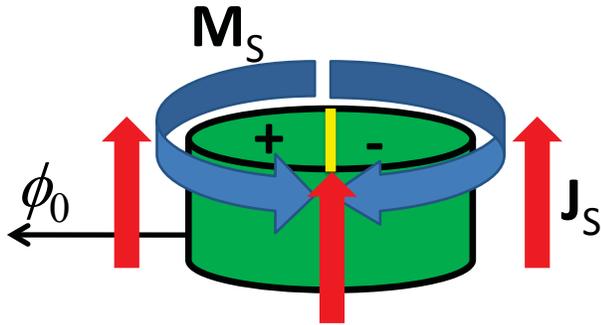}
\caption{(Color online) Sketch of a cylindrical mesa with a surface electric current density ${\bm J}_S$ (red vertical arrows) and a magnetic current density ${\bm M}_S$ (blue horizontal azimuthal arrows).  The $\pm$ signs refer to the signs of the $(11)$ cavity mode, separated by the yellow line, with fixed angle $\phi_0$ indicated by the thin horizontal black arrow.}
\end{figure}\label{fig1}

As long as rectangular mesas were the only radiation sources under study, it has proved difficult to distinguish the results of these radiation sources\cite{KK,Kadowaki2}.  Clearly, the node (or soliton kink) in the center of the mesa assumed by some to be present in the cavity model cannot be the ground state of the system\cite{Tachiki,Tachiki2}, even dynamically, and inhomogeneous, layer-index-dependent kinks are even higher in energy\cite{LinHu,LinHu2,LinHu3}, but the extreme non-linear current-voltage characteristic of each mesa allows for chaotic, nonequilibrium effects to be important, so such complications could potentially play important roles in  the origin of the radiation.  As noted above, additional rectangular cavity modes might form across the length of the sample\cite{Kleiner}.  The only practical ways to distinguish these models in rectangular mesas is by the observations of the angular dependence of the  output power, polarization, and degree of coherence.  As indicated previously\cite{KK}, the antisymmetric cavity model necessarily leads to a maximum in the radiation output power at $\theta=0^{\circ}$, directly above the mesa, whereas the conducting dipole antenna model leads to zero radiation power at $\theta=0^{\circ}$. The situation near $\theta=90^{\circ}$ is much less clear.  The conducting dipole antenna model leads to a substantial output power in that direction, at least in the absence of a superconducting substrate.  However, depending upon the wave vector $k$ of the outgoing radiation, emission from the cavity model can lead either to a vanishing or a non-vanishing output at $\theta=90^{\circ}$, at least for a non-superconducting substrate. Thus, we considered it important to study a different experimental configuration, in which more substantial differences between the output predictions of these two models would be clearly evident.  It is also crucial to consider the effects of the superconducting substrate.
\subsection{C. Cylindrical mesas}
\begin{table}
\begin{tabular}{rrrrrrr}
$m$&$\,\chi_{m1}\>\>$&$\,\chi_{m2}\>\>$&$\,\chi_{m3}\>\>$&$\,\chi_{m4}\>\>$&$\,\chi_{m5}\>\>$&$\,\chi_{m6}\>\>$\\
\hline
\hline
\noalign{\vskip3pt}
0&3.8317&7.0156&10.1735&13.3237&16.4706&19.6159\\
1&1.8412&5.3314&8.5363&11.7061&14.8636&18.0155\\
2&3.0542&6.7061&9.9687&13.1704&16.3475&19.5129\\
3&4.2012&8.0152&11.3459&14.5858&17.7887&20.9725\\
4&5.3175&9.2824&12.6809&15.9641&19.1961&22.4013\\
5&6.4156&10.5205&13.9872&17.3129&20.5755&23.8036\\
6&7.5013&11.7350&15.2682&18.6374&21.9317&25.1840\\
7&8.5778&12.9324&16.5294&19.9405&23.2681&26.5450\\
8&9.6474&14.1155&17.7740&21.2291&24.5872&27.8893\\
9&10.7147&15.2868&19.0046&22.5014&25.8913&29.2186\\
10&11.7699&16.4479&20.2212&23.7607&27.1820&30.5345\\
\hline
\hline
\end{tabular}
\caption{Table of the first six wave vector parameters $\chi_{mp}=k_{mp}a$ for $m=0,\ldots,10$, where $p$ defines the rank ordering of the non-vanishing values of $J'_m(\chi_{mp})=0$, for a cylindrical cavity of radius $a$.}
\end{table}

Here we describe the application of these two models to mesas with cylindrical geometry, which  recently have been studied experimentally\cite{Kadowaki3}.  We note that Hu and Lin numerically studied the radiation from cylindrical mesas suspended in vacuum\cite{HuLin,HuLin2,note}.  In place of Eq. (\ref{krectangle}), the mode frequencies for a cylindrical cavity of radius $a$ with the boundary condition that the azimuthal component of the magnetic field vanishes at the edge, are given by
\begin{eqnarray}
k'_{mp}&=&\chi_{mp}/a,\label{kcylinder}\\
\nu_{mp}&=&\frac{ck'_{mp}}{2\pi n_r}=\frac{c\chi_{mp}}{2\pi an_r},\label{nucylinder}
\end{eqnarray}
where the anharmonic $\chi_{mp}=k'_{mp}a$ values are listed in Table I.  From fits to the fundamental mode on three cylindrical mesas\cite{Kadowaki3}, the choice $n_r\approx\sqrt{18}$ fit the assumption that the fundamental $ac$ Josephson frequency locked onto the cavity $(11)$ mode reasonably well, as it did in fitting the rectangular cavity $(10)$ mode\cite{Ozyuzer,Kadowaki}. Hence, in our numerical calculations, we shall use $n_r=\sqrt{18}$ for the frequencies of interest.

Although a layered superconductor consists of a stack of Josephson junctions that in some circumstances may behave rather independently\cite{KLB}, in our model, we assume that under the application of a $dc$ voltage $V$ across the $N$ layers, all of the layers behave identically.  This assumption was also made by Tachiki {\it et al.},\cite{Tachiki}, and is consistent with the conclusions of Bulaevskii and Koshelev \cite{BK1,BK2,Koshelev}, but is distinctly different from that made by Hu and Lin\cite{LinHu,LinHu2,LinHu3,LinHu4,HuLin,HuLin2}. For the electric dipole antenna model, the frequencies of the emitted radiation are the same as those present in the $ac$ Josephson current, $\nu_n=n\nu_J$, since the integer harmonics are generated by the nonlinear $ac$ Josephson effect.  For the cavity model, however, the frequencies $\nu_{mp}$ of the radiation are the non-vanishing values generated by the $p$ zeroes of the first derivative of the $m$th regular Bessel function $J_m(z)$, which are far from integer multiples of one another, as evidenced in Table I. If a particular cavity mode were to become in resonance with the fundamental $ac$ Josephson frequency $\nu_J$, then higher harmonics of the Josephson current would not be resonant with any other cylindrical cavity mode.  Hypothetically, if  some cavity modes did not require resonance with the Josephson current to radiate, the frequency spectrum of the emitted radiation would be distinctly nonharmonic, as pictured by Hu and Lin\cite{HuLin2}. Hence, if resonance of the fundamental $ac$ Josephson current frequency with a non-uniform cavity mode were crucial for the radiation to occur, one would not expect to observe any higher harmonics in the output power.  Observation of higher harmonics would therefore be {\it prima facie} evidence that a substantial amount of the radiation does not arise from the excitation of cavity modes.

Here we calculate the angle and frequency dependence of the output power for both the cavity and electric dipole models for cylindrical and rectangular mesas.  We show that the experiments on cylindrical mesas presented by Kadowaki {\it et al.} provide evidence that the radiation at the fundamental $ac$ Josephson frequency, consistent with the $(11)$ cylindrical cavity mode, is a mixture obtained from both the electric and magnetic current sources\cite{Kadowaki3}.   To fit the data quantitatively, we must also take account of the dramatic effects of the superconducting substrate. However, the higher frequencies observed in the measured output power are harmonics of the fundamental, so that any mixing with higher cavity modes is very unlikely.  Measurements of the angular dependence of the output power, polarization, and coherence  of the second harmonic of the emission from cylindrical mesas atop superconducting substrates should therefore be fully coherent, have linear polarization, and vanish at $\theta=0^{\circ}$ and $90^{\circ}$. Studies of the angular dependence of the power, polarization, and coherence of the emission at the fundamental frequency may help to separate the two radiation sources, as well. For rectangular mesas, the correct boundary condition of $B_{||}=0$ is difficult to impose precisely in a rectangular geometry, due to the corners. However, with the Love equivalence principles, one can find a closed form expression for the far-field radiation, the results of which will be presented, as well.

The paper is organized as follows.  In Secs. II-VIII, the formation of the radiation emitted from cylindrical cavities is studied theoretically, since the mathematics is the simplest.  In Sec. II, we focus on  the primary radiation source, which we take to be the $ac$ Josephson current.  In Sec. III, we study the spatial fluctuations within the mesa of this primary radiation source. In Sec. IV, we describe how these spatial fluctuations may excite a cavity mode.  In Sec. V, we discuss the radiation from a cavity mode alone.  In Sec. VI, we calculate the output power of the combined primary and excited mode secondary radiation.  In Sec. VII, we calculate and discuss the electric polarization and coherence of this combined radiation.  In Sect. VIII, we discuss and model the effects of superconducting substrates.  In Sec. IX, we present our analogous results for rectangular mesas. We discuss and summarize our results in Sec. X.

\section{II. Primary radiation source}

  As for rectangular mesas, a $dc$ voltage $V$ is applied across the stack of  $N$ superconducting junctions in the BSCCO mesa, which causes the junctions to  exhibit an $ac$ Josephson current, which oscillates at the frequency $\nu_J$ given by Eq. (\ref{nuJ}).
    In addition, the intrinsic non-linearity of the Josephson junctions causes the $ac$ Josephson current to have a large number of harmonics at $\nu_n=n\nu_J$, regardless of the spatial dependence of the Josephson current within each junction. From Love's magnetic equivalence principle, the magnetic field in the cavity generated by the $ac$ Josephson current along $\hat{\bm z}$ according to Amp{\`e}re's law is treated by replacing it with a surface electric current ${\bm J}_S$, and setting the resulting magnetic field in the cavity equal to zero\cite{antenna,antenna2}.  This $ac$ Josephson current has two essential functions in the radiation.  First, it radiates at all of its harmonic frequencies.  Second, the radiation at one of its frequencies may lock onto a cylindrical cavity mode, exciting it, and causing it to radiate, as well.  Hence,  the $ac$ Josephson current is the primary radiation source.    We assume that during emission, all of the $N$ junctions radiate together, so we neglect the layer index of the spatial variation\cite{KLB}, and write within the mesa
   \begin{eqnarray}
   {\bm J}({\bm x}',t)&=&\hat{\bm z}'\sum_{n=1}^{\infty}e^{-in\omega_Jt}[J_n^J+\delta J_n({\bm x}')],\label{Jxt}
   \end{eqnarray}
   where the spatial average $\langle \delta J_n({\bm x}')\rangle$ in the mesa vanishes.  We assume that any time dependence to $\delta J_n({\bm x}')$ is slow with respect to the measurement times, and can be neglected.

    We then may calculate the resulting radiation pattern in the conducting electric dipole antenna model by taking the equivalent electric current source to be the integrated current source placed on the surface from which the radiation occurs, which is the edge of the cylindrical mesa sketched in Fig. 1.  We assume no radiation emanates from the top and bottom surfaces of the cylindrical mesa. This  leads to the surface electric current density, which we write just inside the mesa edge as
\begin{eqnarray}
 {\bm J}_S({\bm x}',t)&=&\frac{a}{2}\eta(z')\delta(\rho'-a){\bm J}({\bm x}',t),\label{JS}
 \end{eqnarray}
where ${\bm J}({\bm x}',t)$ is given by Eq. (\ref{Jxt}), $\eta(z')=\Theta(z')\Theta(h-z')$ is unity over the height $h$ of the mesa, $a$ is the radius  of the cylindrical mesa, and we assume $h/a\ll1$.  In this limit, it suffices to take $\eta(z')\rightarrow h\delta(z')$ when the mesa is suspended in vacuum.  A sketch of the uniform part of the surface electric current density ${\bm J}_S$ at the edge of a cylindrical mesa is given in Fig. 1.

 If the electric surface current density contained only the fundamental $ac$ frequency $\nu_J$, the magnetic
  vector potential ${\bm A}_1({\bm x},t)$ arising from it would be given by

  \begin{eqnarray}
  {\bm A}_1({\bm x},t)&=&\frac{a\mu_0}{8\pi}\int d^3{\bm x}'\hat{\bm z}'\eta(z')\delta(\rho'-a)\frac{e^{i(k_JR-\omega_Jt)}}{R}\nonumber\\
  & &\times[J_1^J+\delta J_1({\bm x}')],\label{A1}
  \end{eqnarray}
  where $R=|{\bm x}-{\bm x}'|$. We note that $\omega_J$ is the same inside and outside the mesa, but  $k_J=\omega_J/c$ outside the mesa must match the wave vector of light in vacuum.  Inside the mesa, $k_J'=n_r\omega_J/c$.   Since ${\bm A}||\hat{\bm z}$ is parallel to the mesa edge, this wave vector automatically satisfies the Maxwell boundary condition that the tangential components of ${\bm E}$ are preserved across the boundary. Generalizing this standard expression to the full set of harmonic $ac$ Josephson frequencies, we have
  \begin{eqnarray}
  {\bm A}({\bm x},t)&=&\frac{a\mu_0}{8\pi}\sum_{n=1}^{\infty}\int d^3{\bm x}'\hat{\bm z}'\eta(z')\delta(\rho'-a)\frac{e^{in(k_JR-\omega_Jt)}}{R}\nonumber\\
  & &\qquad\times[J_n^J+\delta J_n({\bm x}')].\label{An}\nonumber\\
  \end{eqnarray}
  We write
  \begin{eqnarray}
  {\bm A}({\bm x},t)&=&{\bm A}_J({\bm x},t)+{\bm A}_{\delta J}({\bm x},t),\label{AJplusAdeltaJ}
   \end{eqnarray}
   separating it into its contributions from the spatially uniform $J_n^J$ and  inhomogeneous $\delta J_n^J({\bm x}')$ source amplitudes. In Sec. III, we present a theory for $\delta J({\bm x}')$, and use it to evaluate ${\bm A}_{\delta J}({\bm x},t)$.

  In the far field, or radiation zone, $r/a\gg1$, and we use the standard approximation
  \begin{eqnarray}
  \frac{e^{ikR}}{R}&\rightarrow&\frac{e^{ikr}}{r}e^{-i{\bm k}\cdot{\bm x}'},
  \end{eqnarray}
  where $|{\bm x}|=r$ in spherical coordinates\cite{antenna,antenna2,Jackson}.  In the radiation zone, $\hat{\bm z}'\rightarrow-\hat{\bm \theta}\sin\theta$ and ${\bm k}\cdot{\bm x}'=k\rho'\sin\theta\cos(\phi-\phi')$.
  Taking $h/a\ll1$, so that $\eta(z')\rightarrow h\delta(z')$ when the mesa is suspended in vacuum (in a Gedanken experiment), the integral in Eq. (\ref{An}) is then readily found to be
  \begin{eqnarray}
  {\bm A}_J({\bm x},t)&{\rightarrow\atop r/a\gg1}&-\hat{\bm \theta}\frac{v\mu_0\sin\theta}{4\pi r}\sum_{n=1}^{\infty}e^{in(k_Jr-\omega_Jt)}J_n^J\nonumber\\
  & &\qquad\times J_0(nk_{\theta})S_n^J(\theta),\label{AJ}
  \end{eqnarray}
  where $v=\pi a^2h$ is the volume of the mesa, $k_{\theta}=k_Ja\sin\theta$, and $S_n^J(\theta)=1$ in the absence of a substrate.
Then, it is elementary to obtain ${\bm E}_{\bm A}({\bm x},t)$ from Maxwell's equation, ${\bm E}_{\bm A}=-\frac{\partial {\bm A}}{\partial t}$, since there is no electric potential outside the mesa. Similarly, ${\bm H}_{\bm A}=\frac{1}{\mu_0}{\bm\nabla}\times{\bm A}$. In the radiation zone, the contribution ${\bm E}_{{\bm A}_J}$ from ${\bm A}_J$ is given by
\begin{eqnarray}
{\bm E}_{{\bm A}_J}({\bm x},t)&{\rightarrow\atop r/a\gg1}&-\frac{i\hat{\bm \theta}\sin\theta v\mu_0}{4\pi r}\sum_{n=1}^{\infty}e^{in(k_Jr-\omega_Jt)}\nonumber\\
& &\qquad\times nJ_n^J\omega_JJ_0(nk_{\theta})S_n^J(\theta).
  \end{eqnarray}

  We note that this implies the radiation from the surface electric current source is linearly polarized, with its polarization direction parallel to $\hat{\bm\theta}$.  So far, however, $k_J=\omega_J/c$ is only determined by the applied $V/N$, and is unspecified with respect to the mesa dimensions.  When the fundamental $ac$ Josephson mode with frequency $\nu_J$ locks onto the $(mp)$ cavity mode, the resulting $k_J'$ and $k_J$ values are given by $k_J'a=\omega_Jn_r/c=\chi_{mp}$ inside and $k_Ja=\omega_J/c=\chi_{mp}/n_r$ outside the mesa, in order to properly describe the electromagnetic propagation in the vacuum.

  \begin{figure}
\includegraphics[width=0.235\textwidth]{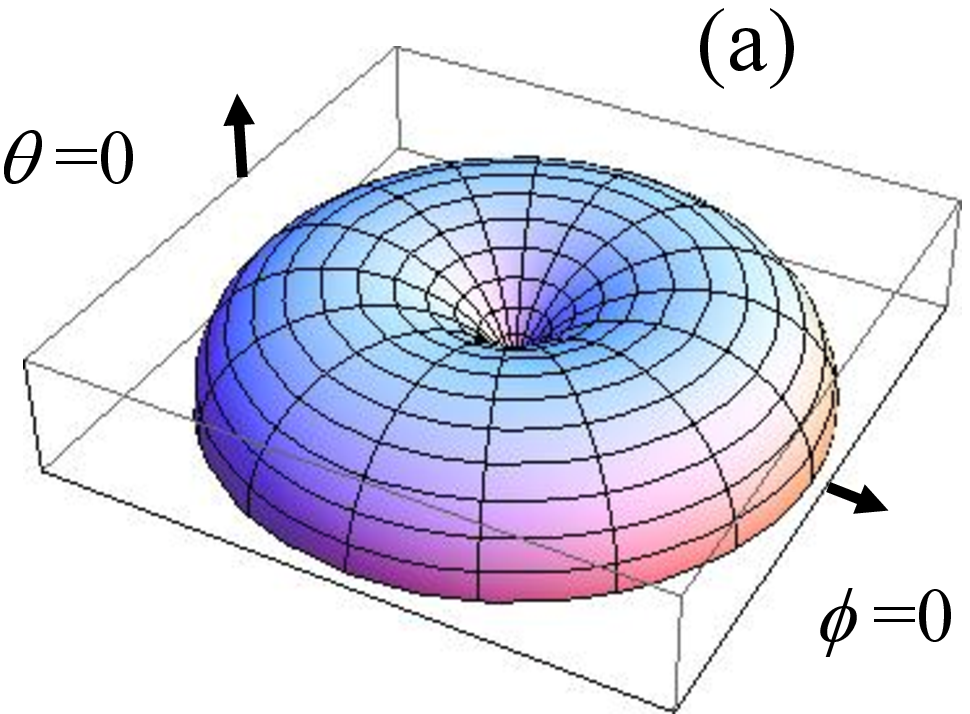}\hskip5pt
\includegraphics[width=0.235\textwidth]{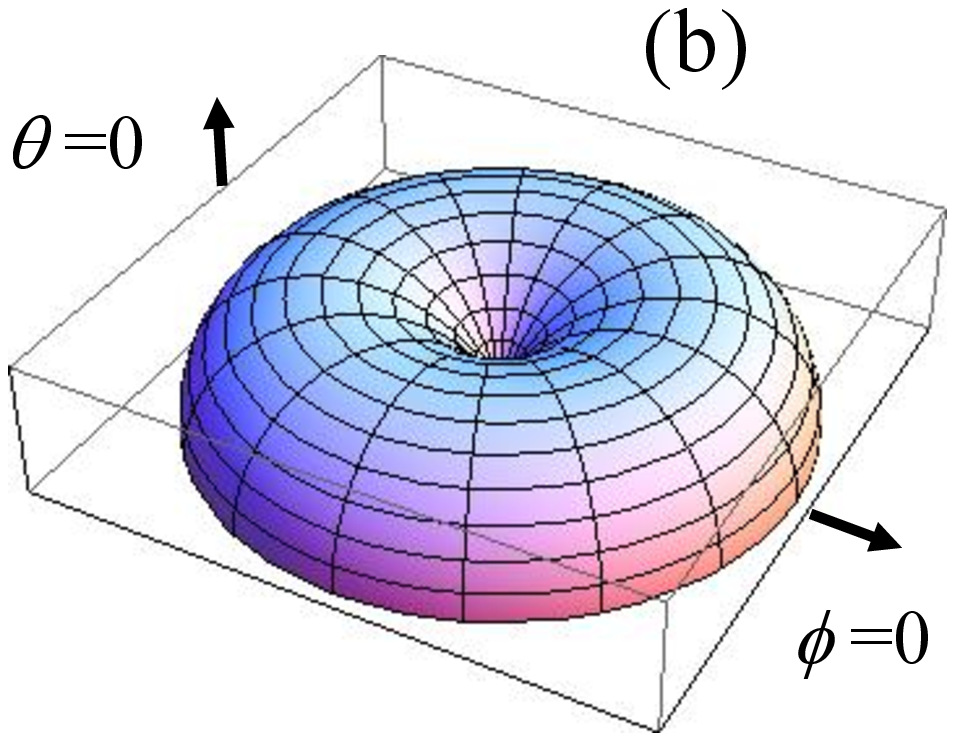}
\caption{(Color online) Plots of the radiation intensity in arbitrary units emitted from the uniform $J_n^J$ part of the $ac$ Josephson current when a cylindrical mesa is suspended in vacuum. (a) The radiation at the fundamental $n=1$ mode. (b)  The radiation at the $n=2$ second harmonic mode.   }\label{fig2}
\end{figure}

  The radiation-zone intensity $I(\theta,\phi)$ for the conducting dipole model is given by the differential power per unit solid angle by $dP/d\Omega=\frac{1}{2}{\rm Re}[r^2\hat{\bm r}\cdot\overline{{\bm E}\times{\bm H}^{*}}]$\cite{Jackson}, where $\overline{\cdots}$ is a time average.  The contribution from the uniform $J_n^J$ source in the radiation zone is
\begin{eqnarray}
\frac{dP_J}{d\Omega}&{\rightarrow\atop r/a\gg1}&\sin^2\theta\sum_{n}|B_n(\theta)J_0(nk_{\theta})|^2,\label{Pn}
\end{eqnarray}
where $B_n(\theta)=J^J_nvS_n^J(\theta)nk_J\sqrt{Z_0}/(4\sqrt{2}\pi)$, and $Z_0=\sqrt{\mu_0/\epsilon_0}$ is the vacuum impedance.
The radiation patterns expected with $k_Ja=1.8412/n_r=\chi_{11}/n_r$ from the uniform part of the conducting dipole model alone at the fundamental $(n=1$) and second harmonic $(n=2)$ when the mesa is suspended in vacuum are pictured in Figs. 2(a) and 2(b), respectively.  These intensity patterns are nearly indistinguisable, except for the emission frequency, due to the long wavelength in vacuum relative to the mesa size.  The intensities vanish at $\theta=0^{\circ}$, due to the geometric factor $\sin\theta$ in the radiation-zone spherical coordinate representation of $\hat{\bm z}'$, are independent of $\phi$, and  are non-vanishing at $\theta=90^{\circ}$.  When a cylindrical mesa is suspended in vacuum, $I(\theta,\phi)$ for the fundamental $n=1$ and second harmonic $n=2$ modes each exhibit a maximum at $\theta=90^{\circ}$,  as for the ordinary dipole model based upon a quasi-one-dimensional wire or thin rod-shaped source.
\section{III. Spatial fluctuations of the $ac$ Josephson current}

If the $ac$ Josephson current were independent of position within the mesa, this could only couple to a uniform electric field, which would not radiate.  Hence, the non-uniform part $\delta J({\bm x}')$ of  the $ac$ Josephson current must provide the coupling.  There may be a number of sources for this non-uniformity, such as defects and thermal fluctuations, which may be aggravated by inhomogeneous heating effects\cite{Kleiner}.  Here we assume that the primary source of the inhomogeneities is thermal fluctuations, but our model can also be employed to approximate smooth stoichiometry variations, as well\cite{inhomog1,inhomog2}.  In the samples used in the experiments\cite{Kadowaki3}, the BSCCO crystals were prepared slowly over a very limited central region in gold-plated elliptical heating ovens, with highly controlled heating,  producing samples that are probably much more uniform chemically (although still highly non-stoichiometric) than in those studies\cite{Kadowaki4,Kadowaki5}.

 Assuming the magnitude of the fluctuations in the spatial dependence of the $ac$ Josephson current is small within the mesa with respect to its average value, the free energy of these fluctuations may be written as
\begin{eqnarray}
{\cal F}_f&\propto&\sum_{n=1}^{\infty}\int d^2{\bm r}'[(\delta J_n)^2+(\xi'_n)^2({\bm\nabla}'\delta J_n)^2],\label{Ff}
\end{eqnarray}
where $\xi'_n(T)$ is a temperature-dependent characteristic length over which the spatial fluctuations in $\delta J_n$ occur, and $n$  describes the intralayer spatial variations associated with the $n$th harmonic of the $ac$ Josephson frequency, as noted in Eq. (\ref{Jxt}).  ${\cal F}_f$ is a minimum when
\begin{eqnarray}
-(\xi'_n)^2({\bm\nabla}')^2\delta J_n+\delta J_n&=&0,
\end{eqnarray}
which  is written in cylindrical coordinates.  The full solution to this equation is elementary, and may be written as
\begin{eqnarray}
\delta J_n(\rho',\phi')&=&\sum_{m=0}^{\infty}C^{(n)}_{m}J_m(\rho'/\xi'_n)\cos[m(\phi'-\phi_{0n})],\label{deltaJ}\nonumber\\
\end{eqnarray}
where the $\phi_{0n}$ and  $C^{(n)}_{m}$ are free parameters.
Of course, we require that the $ac$ Josephson current vanishes outside the mesa ($\rho'>a$), but since the uniform part $J_n^J$ is discontinuous at the mesa edge, $\delta J_n$ may also be discontinuous at $\rho'=a$.  Instead, we assume that thermal fluctuations are weak near the mesa edge, so that we take
\begin{eqnarray}
\frac{\partial\delta J_n}{\partial \rho'}\Bigl|_{\rho'=a}&=&0.
\end{eqnarray}
With this von Neumann boundary condition, $\delta J_n({\bm x}')$ may be written as
\begin{eqnarray}
\delta J_n({\bm x}')&=&\sum_{p=1,m=0}^{\infty}C^{(n)}_{mp}J_m(k'_{mp}\rho')\nonumber\\
& &\qquad\times\cos[m(\phi'-\phi_{0n})],\label{deltaJform}
\end{eqnarray}
where $\chi_{mp}=k'_{mp}a$ is the  $p$th non-vanishing value of $dJ_m(x)/dx=0$, and we write $k'_{mp}=1/\xi'_{mp}$.  The   $\chi_{mp}$ values for $m=0,\ldots,10$ and $p=1,\ldots,6$ are give in Table I.  Since $\int_0^a xdxJ_0(k_{0p}x)=(a/k'_{0p})J_1(k'_{0p}a)=0$ for all allowed $p$ values, the spatial average of $\delta J_n({\bm x}')$ vanishes.  We note that $k'_{mp}$ represents the set of $1/\xi'_n$ values that satisfy our  assumed boundary condition.  In addition, since $J_n^J$ and $\delta J_n({\bm x}')$ are real, all of the $C_{mp}^{(n)}$ are real.

In the Appendix, we calculate the electric field ${\bm E}_{{\bm A}_{\delta J}}({\bm x},t)$ due to this inhomogeneous $ac$ Josephson current distribution and the resulting combined ${\bm E}_{{\bm A}}({\bm x},t)={\bm E}_{{\bm A}_{J}}({\bm x},t)+{\bm E}_{{\bm A}_{\delta J}}({\bm x},t)$, and discuss the complications that arise.

\section{IV. Excitation of a cavity mode}

We next consider the cavity model of the short cylindrical mesa.    Since a voltage is applied across the height of the mesa, the magnetic vector potential  $A_z$ satisfies ${\bm\nabla}'^2A_z-\mu\epsilon\frac{\partial^2A_z}{\partial t^2}=-\mu J_z$ inside the mesa, where $\mu$ and $\epsilon$ are the magnetic permeability  and dielectric constant inside it, and $n_r=\sqrt{\mu\epsilon}\approx\sqrt{\epsilon}$.
In the absence of an $ac$ Josephson current, $J_z=0$,  the homogeneous solution $A^{(h)}_{z}$ is
\begin{eqnarray}
A^{(h)}_{z}({\bm x}',t)&=&\sum_{m=0,p=1}^{\infty}B_{mp}e^{-i\omega_{mp}t}J_m(k'_{mp}\rho')\nonumber\\
& &\qquad\times\cos[m(\phi'-\phi'_{0})],\label{Azh}
\end{eqnarray}
where   $\phi_0'$ and the $B_{mp}$ are arbitrary constants, and we neglected spatial variations along the $z$ axis. As part of Love's magnetic equivalence principle, Eq. (\ref{Azh}) is subject to the condition $-\frac{\partial A_z}{\mu\partial\rho'}\Bigr|_{\rho'=a}=H_{\phi}\Bigr|_{\rho'=a}=0$ \cite{antenna,antenna2}. The consequence of this requirement and the related feature of setting ${\bm H}({\bm x}',t)=0$ inside the cavity, is to replace the internal magnetic field by a surface electric current density ${\bm J}_S$. Hence, this boundary condition is precise, and is not an assumption, as we made for $\delta J_n$.  This condition then leads to the same wave vectors $k'_{mp}=\chi_{mp}/a$ as for $\delta J_n$.
Adding the  spatially inhomogeneous part  $\delta J_n({\bm x}')$ to the wave equation for $A_z$, we have
\begin{eqnarray}
{\bm\nabla}'^2A_z-\mu\epsilon\frac{\partial^2A_z}{\partial t^2}&=&-\mu\sum_{m=0; p,n=1}^{\infty}e^{-in\omega_Jt}C^{(n)}_{mp}\nonumber\\
& &\times J_m(k'_{mp}\rho')\cos[m(\phi'-\phi_{0n})],\label{fundamental}\nonumber\\
\end{eqnarray}
which can be solved with the particular resonance solution form
\begin{eqnarray}
A^{(p)}_{z}({\bm x}',t)&=&\sum_{m=0;n,p=1}^{\infty}e^{-in\omega_Jt}[A^{(n)}_{mp}(t)\delta_{m,m_0}\delta_{p,p_0}\nonumber\\
 & &+A^{(n)}_{mp}(1-\delta_{m,m_0}\delta_{p,p_0})]J_m(k'_{mp}\rho')\nonumber\\
 & &\qquad\times\cos[m(\phi'-\phi_{0n})],\label{Ap}
\end{eqnarray}
provided that
\begin{eqnarray}
A^{(n)}_{m_0p_0}(t)&=&t\frac{iC^{(n)}_{m_0p_0}}{2n\epsilon\omega_J}\qquad\>\>\>\>\>\>{\rm for}\>(mp)=(m_0p_0),\label{Am0p0oft}\\
\omega_{m_0p_0}^2&=&n^2\omega_J^2\qquad\qquad\>\>\>{\rm for}\>(mp)=(m_0p_0),\label{omegam0p0}\\
A^{(n)}_{mp}&=&\frac{C^{(n)}_{mp}}{\epsilon(\omega_{mp}^2-n^2\omega_J^2)}\>{\rm for}\>(mp)\ne(m_0p_0),\label{Anmp}
\end{eqnarray}
where we have set $k'_{m_0p_0}=\omega_{m_0p_0}n_r/c$.
The spatially constant part of $J_J({\bm x}',t)$ cannot excite any cavity modes. Although  Eqs. (\ref{Am0p0oft}) can be solved for general $(mp)$,  Eq. (\ref{omegam0p0}) can only hold for   one value of $(mp)$, which we denote $(m_0p_0)$. Of course, if no $(m_0p_0)$ satisfies Eq. (\ref{omegam0p0}), then none of the cavity modes will be in resonance with any of the modes of the inhomogeneous $ac$ Josephson current. Hence, we conclude that the non-linear, spatially inhomogeneous Josephson current can  excite at most one electromagnetic cavity mode, and the other modes are off resonance, with Lorentzian lineshapes as suggested by Eq. (\ref{Anmp}).  Of course,  effects such as a built-in $z$-dependence to the cylinder radius $a$ during mesa fabrication will give finite widths $\propto 1/Q_{mp}$ to the cavity modes. Note that the $t$ in the solution for the excited $(m_0p_0)$ mode causes the amplitude $A^{(n)}_{m_0p_0}(t)$ of that mode to grow linearly in time, which continues until saturation is achieved, and radiation occurs.  During radiation, the amplitude of the resonant frequency is approximately constant in time, which we take to have the finite value $A^{(n)}_{m_0p_0}(\infty)\equiv\lim_{t\rightarrow\infty}A^{(n)}_{m_0p_0}(t)$.  We note from Eq. (\ref{Am0p0oft}), that for small times, the resonant amplitude $A_{m_0p_0}^{(n)}(t)$ is pure imaginary for a real $\epsilon$ in the frequency range of interest, so we assume that it stays pure imaginary after saturation, and write
\begin{eqnarray}
A^{(n)}_{m_0p_0}(\infty)&=&it_{\rm eff}\frac{C_{m_0p_0}^{(n)}}{2n\epsilon\omega_J},\label{Am0p0infty}
\end{eqnarray}
where $t_{\rm eff}$ is the effective time to reach saturation.

Since the lowest energy cavity mode is the $(mp)=(11)$ mode, if all of the junctions are involved in the radiation, it will be the easiest mode to excite by experimentally increasing $V$.  If the upturn in the current-voltage ($I/V$) characteristic occurs before this mode can be excited, then no resonance of the $ac$ Josephson fundamental mode including  all $N$ junctions with any cavity mode can be achieved.
\begin{figure}
\includegraphics[width=0.48\textwidth]{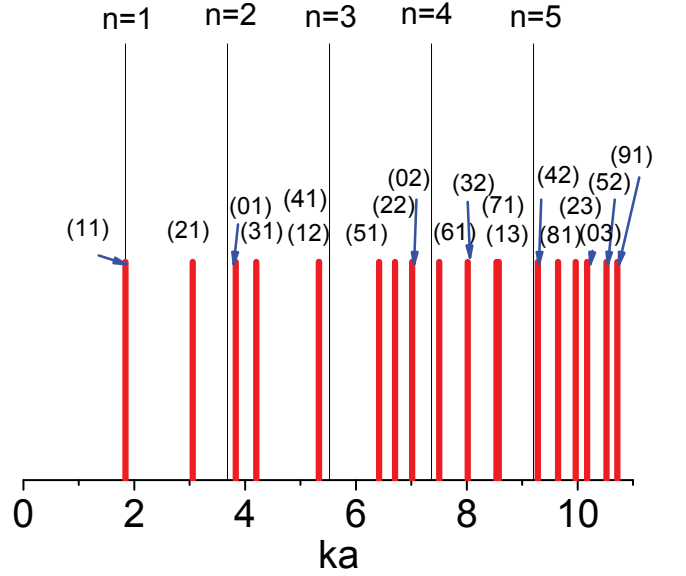}
\caption{(Color online) Sketch of the lowest five wave vectors of the $ac$ Josephson current harmonics (long thin black) and the lowest 19 cylindrical cavity modes (short thick  red), assuming the fundamental $(n=1)$ $ac$ Josephson frequency locks onto the lowest energy $(11)$ cavity mode frequency. The $(mn)$ indices of the cavity modes are indicated.}\label{fig3}
\end{figure}
\section{V. Radiation from a cylindrical cavity mode}

Now,  we
 focus our attention on the angular dependence of the radiation, assuming that a particular cavity mode $(mp)=(m_0p_0)$ is involved in the radiation.  We assume that after equilibrium has been achieved, the magnetic vector potential from which the electric field in the cavity can be calculated is given by the particular form, Eq. (\ref{Ap}).  Then, the magnetic surface current density ${\bm M}_S({\bm x}',t)=-\hat{\bm n}\times {\bm E}^{(p)}({\bm x}',t)$, evaluated at the sample edge,\cite{antenna,antenna2} where $\hat{\bm n}=\hat{\bm\rho}'$,
 \begin{eqnarray}
{\bm M}_S({\bm x}',t)&=&-\hat{\phi}'\frac{a}{2}\eta(z')\delta(\rho'-a)\frac{\partial A^{(p)}_z({\bm x}',t)}{\partial t},\label{MS}
\end{eqnarray}
where $\eta(z')$ is defined following Eq. (\ref{JS}).

\begin{figure}
\includegraphics[width=0.235\textwidth]{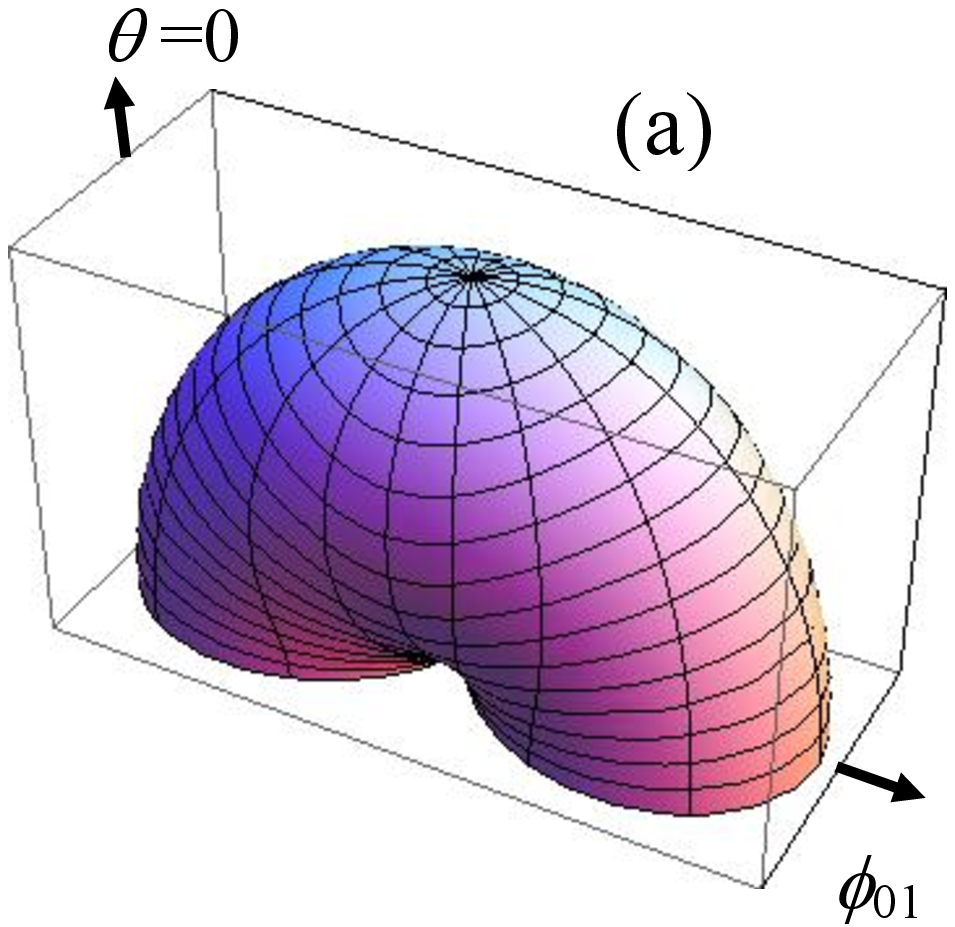}\hskip5pt
\includegraphics[width=0.235\textwidth]{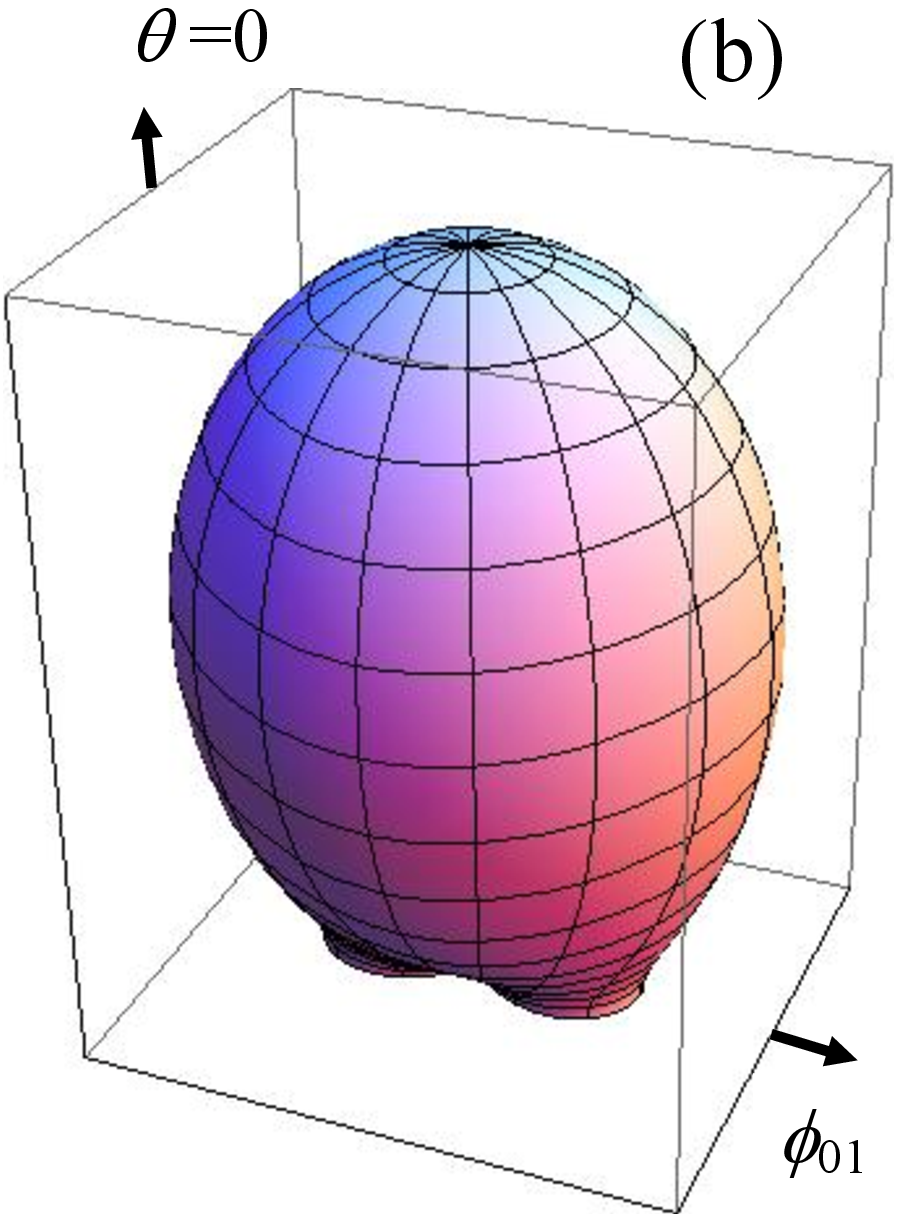}
\caption{(color online) Plots of the radiation intensity in arbitrary units emitted from cavity modes when the mesa is suspended in vacuum. (a) The cavity $(11)$ mode with $k_{11}a=1.8412/n_r$. (b) The cavity $(12)$ mode with $k_{12}a=5.3314/n_r$. }\label{fig4}
\end{figure}

\begin{figure}
\includegraphics[width=0.235\textwidth]{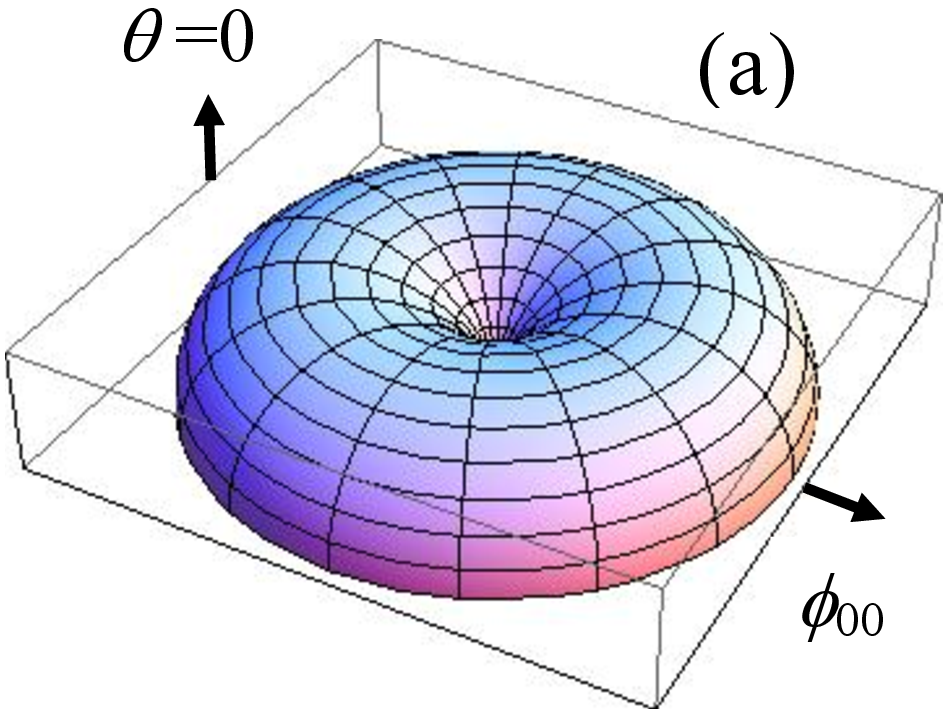}\hskip5pt
\includegraphics[width=0.235\textwidth]{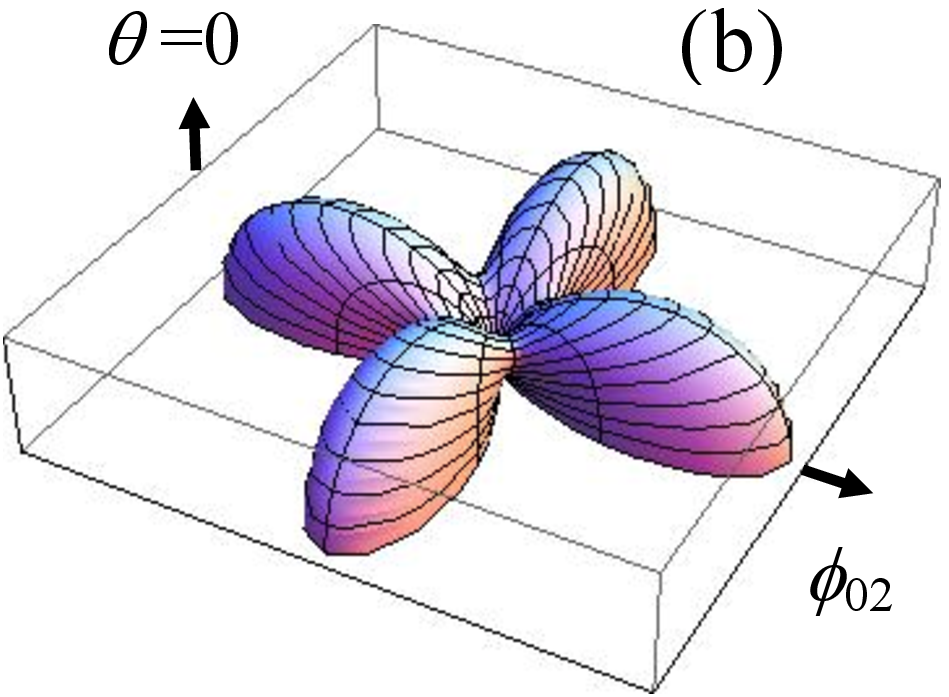}
\caption{(Color online) Plots of the radiation intensity in arbitrary units emitted from cavity modes when the mesa is suspended in vacuum. (a) The cavity (01) mode with $k_{01}a=3.8317/n_r$. (b) The cavity $(21)$ mode with $k_{21}a=3.0542/n_r$.}\label{fig5}
\end{figure}

For a single frequency $\omega$ and wave vector $k$, one obtains the electric vector potential ${\bm F}({\bm x},t)$ outside the sample by the standard integral
\begin{eqnarray}
{\bm F}({\bm x},t)&=&\frac{\epsilon_0}{4\pi}\int d^3{\bm x}'{\bm M}_S({\bm x}',t)e^{ikR}/R,\label{F}
 \end{eqnarray}
 where  ${\bm M}_S({\bm x}',t)$ is the surface magnetic current density at the frequency $\omega=ck$\cite{antenna,antenna2,Jackson}.

  The generalized ${\bm F}({\bm x},t)$ is then obtained from Eq. (\ref{F}) by replacing $\omega$ with $\omega_{m_0p_0}=n\omega_J$, and the details are given in the Appendix.
   Using the Schelkunoff expression for the electric field, ${\bm E}_{\bm F}=-\frac{1}{\epsilon_0}{\bm\nabla}\times{\bm F}$, we obtain for the resonant contribution to ${\bm E}_{\bm F}$ in the radiation zone,
  \begin{eqnarray}
  {\bm E}^{\rm res}_{\bm F}({\bm x},t)&{\rightarrow\atop r/a\gg1}&-(-i)^{m}G^{(n)}_{mp}\frac{e^{i(k_{mp}r-\omega_{mp}t)}}{4\pi r}\nonumber\\
  & &\times\bigl\{\hat{\bm\phi}\cos\theta\sin[m(\phi-\phi_{0n})]J_m^{+}(k_{mp}^{\theta})\nonumber\\
  & &+\hat{\bm\theta}\cos[m(\phi-\phi_{0n})]J_m^{-}(k_{mp}^{\theta})\bigr\}\Bigr|_{{p=p_0}\atop{m=m_0}}.\label{EF}
 \end{eqnarray}

The magnetic field ${\bm H}_{\bm F}$ at the frequency of interest may be obtained from $-\frac{\partial{\bm H}_{\bm F}}{\partial t}={\bm\nabla}\times{\bm E}_{\bm F}$.
We note that the overall factor $-(-i)^{m}$ in ${\bm E}^{\rm res}_{\bm F}$ plays a very important role when ${\bm E}_{\bm A}$ is added to ${\bm E}^{\rm res}_{\bm F}$, as discussed in the following.

 We then calculate  the output  power of the $(m_0p_0)$ cavity mode in the radiation zone, which is
 averaged over $t$ after the resonance has saturated.  We note that some or all of the $\phi_{0n}$ might vary with experimental run, in which case an average of the output power over all possible $\phi_{0n}$ values might be warranted.  Assuming the $(m_0p_0)$ mode of the cavity model of a cylindrical mesa can be excited sufficiently in order to radiate, its differential power per unit solid angle  in the radiation zone at fixed $\phi_{0n}$ is then
\begin{eqnarray}
\frac{dP_{mp}}{d\Omega}&{\rightarrow\atop r/a\gg1}&\frac{\sqrt{\epsilon}(G^{(n)}_{mp})^2}{32\pi^2}\Bigl[\Bigl(\cos[m(\phi-\phi_{0n})]J_m^{-}(k_{mp}^{\theta})\Bigr)^2\nonumber\\
& &+\Bigl(\cos\theta\sin[m(\phi-\phi_{0n})]J_m^{+}(k_{mp}^{\theta})\Bigr)^2\Bigr]\biggr|_{{p=p_0}\atop{m=m_0}}.\nonumber\\
\label{Pmp}
\end{eqnarray}

  In the radiation zone, spherical plots of the radiation patterns for the  four TM$^z_{mp}$ cavity modes with the lowest energies and with $\phi_{0n}$ fixed are shown for a cylindrical mesa suspended in vacuum in  Figs.  4-5.  For fixed $\phi_{0n}$, the radiation patterns all exhibit $C_{2mv}$ point group symmetry, invariant under rotations of $\pi/{m}$ about the $\theta=0$ axis, and exhibiting $2m$ mirror planes. For $m=0$, we may denote the point group symmetry as $C_{\infty v}$, as it is rotationally invariant about $\theta=0^{\circ}$, with a limitless number of mirror planes.  We note that the far-field radiation from the $m=1$ cavity modes, two of which are  pictured in Fig. 4,  exhibit absolute maxima at $\theta=0$, suggestive of a relationship with the experiments, and they exhibit a  substantial azimuthal anisotropy, especially for $\theta\approx90^{\circ}$, where the output from the experiments is small or may even vanish\cite{Kadowaki3}, unlike the azimuthal average of the patterns shown in Figs. 4 and 5.  The output power when the mesa is suspended in vacuum from all other cavity modes, such as the $(01)$ and $(21)$ modes pictured in Fig. 5 and those of the not-pictured $(31)$ and $(41)$ modes, vanish at $\theta=0$, unlike the experiments.  All cavity modes emitted from mesas suspended in vacuum have large outputs for  $\theta=90^{\circ}$.  The output power of the cavity $(01)$ mode pictured in Fig. 5(a)  is independent of $\phi$, as is evident from Eq. (\ref{Pmp}), and is very similar to that of the $ac$ Josephson $n=2$ mode pictured in Fig. 2(b).  We remark that Hu and Lin presented very similar figures for the same four modes that we showed in Figs. 4-5  recently\cite{HuLin2,note}.

\section{VI. Combined primary and secondary radiation}
After equilibrium has been achieved, we assume the fundamental $n=1$ $ac$ Josephson mode has locked onto the cavity $(m_0p_0)=(11)$ mode, with the wave vector $k_{11}a=1.8412/n_r$, as seen in experiments on three cylindrical mesas\cite{Kadowaki3}.  The combined electric field  will then have the components
\begin{eqnarray}
E_{\theta}&=& i[G^{(1)}_{11}\cos(\phi-\phi_{01})J_1^{-}(k_{\theta})-\overline{B}_1\sin\theta J_0(k_{\theta})]\nonumber\\
& &\times\frac{e^{i(k_Jr-\omega_Jt)}}{4\pi r},\label{Etheta1}\\
E_{\phi}&=&iG^{(1)}_{11}\cos\theta\sin(\phi-\phi_{01})J_1^{+}(k_{\theta})\frac{e^{i(k_Jr-\omega_Jt)}}{4\pi r},\label{Ephi1}\\
\overline{B}_n&=&nv\mu_0\omega_JJ_n^JS_n^J(\theta),\label{Bn}
\end{eqnarray}
Because $-(-i)^{m}=i$ for $m=1$, the components to $E_{\theta}$ from the spatially homogeneous $J_n^J$ contribution to ${\bm E}_{\bm A}$ and from ${\bm E}^{\rm res}_{\bm F}$ are either in phase or out of phase by $\pi$, depending upon $\phi_{01}$. However, when the inhomogeneous contributions $\delta J_n^J({\bm x}')$ are included in ${\bm E}_{\bm A}$, as shown in Eq. (\ref{EAfull}),  the  contribution proportional to $\cos[m(\phi-\phi_{0n})]$ to the full ${\bm E}_{\bm A}$ is proportional to $(-i)^{m+1}$, so that this contribution would be out of phase by $\pm\pi/2$ with the two terms included in Eq. (\ref{Etheta1}). If significant in size, such terms would have an important, experimentally observable effect,  both upon the angular dependence  and the polarization of the radiation.

 The combined output power of the fundamental $n=1$, spatially homogeneous $ac$ Josephson surface current density mode locked onto the cavity $(11)$ mode at $\omega_J=\omega_{11}$ is given in the radiation zone by
 \begin{eqnarray}
\frac{dP_{11}}{d\Omega}&{\rightarrow\atop r/a\gg1}&\frac{\sqrt{\epsilon}}{32\pi^2}\Bigl[\Bigl(G^{(1)}_{11}\cos\theta\sin(\phi-\phi_{01})J_1^{+}(k_{\theta})\Bigr)^2\nonumber\\
& &+\Bigl(G^{(1)}_{11}\cos(\phi-\phi_{01})J_1^{-}(k_{\theta})\nonumber\\
& &\qquad-\overline{B}_1\sin\theta J_0(k_{\theta})\Bigr)^2\Bigr].
\label{P11}
\end{eqnarray}
   Note that since the contributions to $E_{\theta}$ from electric and magnetic surface current densities are either in phase or $\pi$ out of phase for the $(11)$ cavity mode locked onto the $n=1$ dipole mode, the fully coherent output power is asymmetric in $\phi$, breaking the $C_{2v}$ point group symmetry to $C_1$.  There remains only a single mirror plane, which is normal to the substrate and contains $\phi_{01}$.  A plot of the combined output power from the  $(11)$ and $n=1$ modes for $\overline{B}_1/G^{(1)}_{11}=5$ when $\phi_{01}$ is fixed and the mesa is suspended in vacuum is shown in Fig. 6(a).  This figure is in rather good agreement with experiment, except for the behavior near to $\theta=90^{\circ}$, where the experimental values are very small, and may even vanish\cite{Kadowaki2}.   There is a point kink, a point where the output power is much smaller than in the surrounding solid angle, which lies along the $\phi_{01}$ line.  Detector scans across the cylindrical mesa top that  pass this point are only infinitesimally probable, and hence may record a significant output power when the detector passes nearby but not directly over the position of the point kink.  However, we note that this simple output form arose from the neglect of the spatially inhomogeneous $ac$ Josephson current $\delta J_n^J({\bm s}')$ source to ${\bm E}_{\bm A}$, as discussed in Sec. III. Inclusion of the contribution proportional to $\cos(\phi-\phi_{01})$ would be $\pm\pi/2$ out of phase with the two components in Eq. (\ref{Etheta1}), and would tend to remove the point kink present in Eq. (\ref{P11}) and shown in Fig. 6(a), and weaken the overall degree of coherence.  Higher azimuthally anisotropic terms proportional to $\cos[m(\phi-\phi_{01})]$ would also be present.  Hence, if it can be established both that $\phi_{01}$ is not random, and second that such a kink in the angle dependence of the output power is absent, then that would provide evidence for a second role for the $\delta J_n^J({\bm x}')$ inhomogeneous electric current source, besides its excitation of the appropriate cavity modes.   Note that the output power is finite at both $\theta=0^{\circ}$ and $\theta=90^{\circ}$.

   In Fig. 6(b), a plot of the output power of the incoherently combined $n=1$ dipole and $(11)$ cavity modes with $\phi_{01}$ random and $\overline{B}_1/G^{(1)}_{11}=7$ is shown.  This output is taken by averaging Eq. (\ref{Pmp}) over $\phi_{01}$, which can be done by inspection.  But it has a drastic effect on the output point group symmetry, which is now $C_{\infty v}$, exhibiting rotational invariance about $\theta=0^{\circ}$. As seen in Sec. VIII, it also has a strong effect on the polarization.
\begin{figure}
\includegraphics[width=0.235\textwidth]{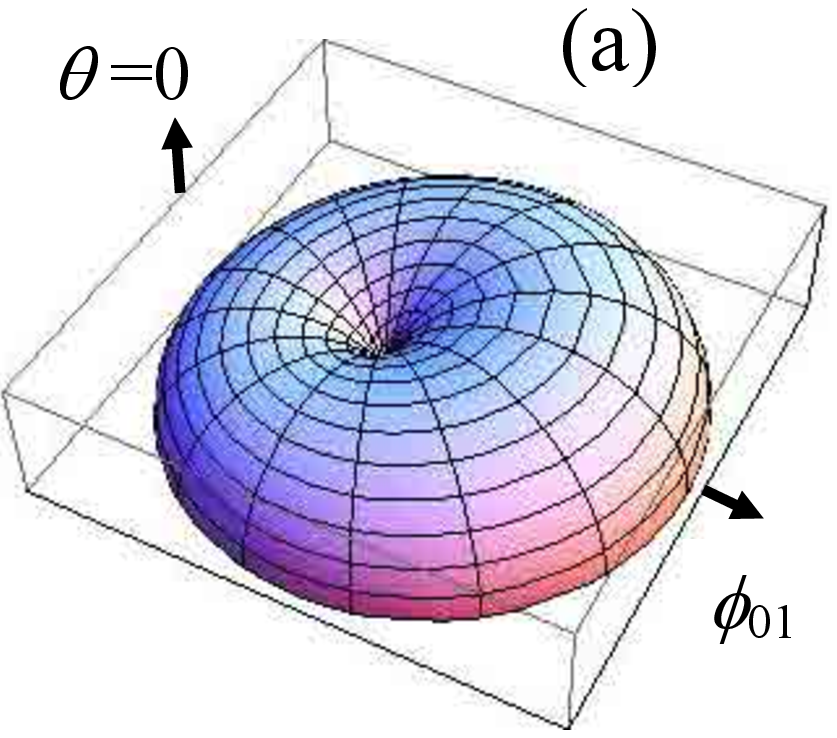}\hskip5pt
\includegraphics[width=0.235\textwidth]{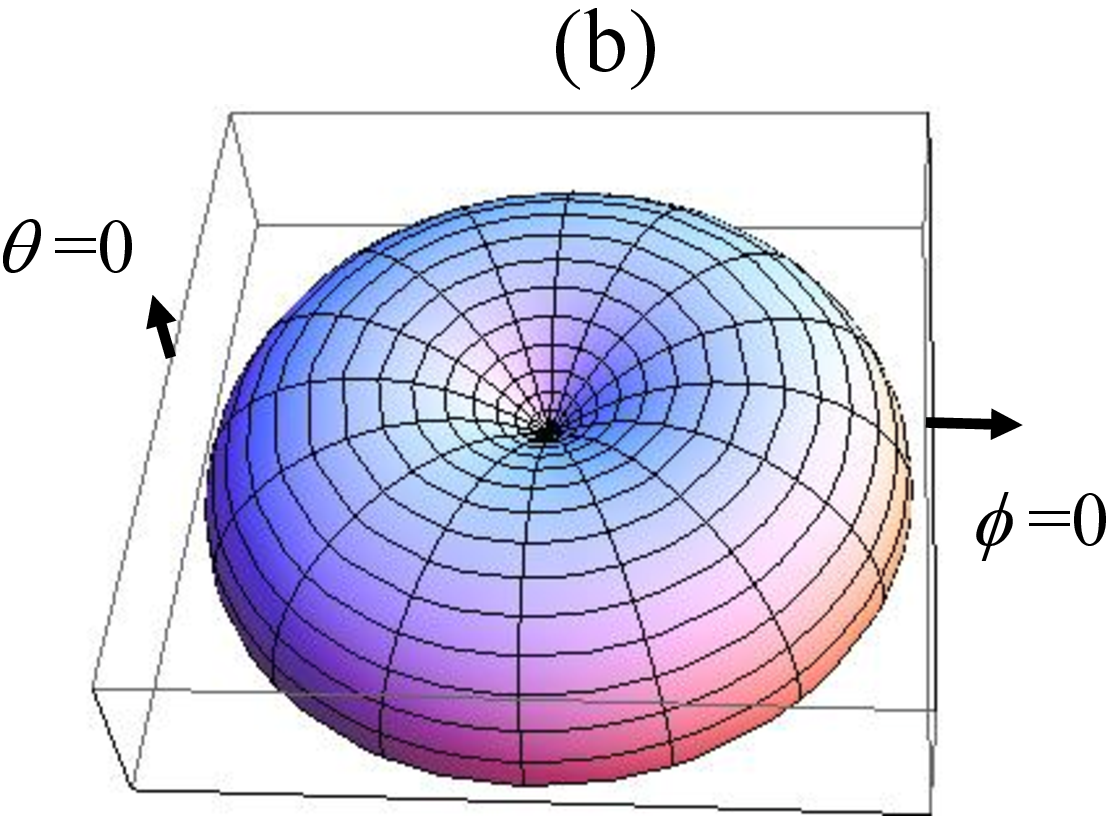}
\caption{(Color online) Plots of  the  radiation power emitted from the mixed $n=1$ $ac$ Josephson and $(11)$ cavity (11) mode with $\phi_{01}$ fixed and  $k_{11}a=1.8412/n_r$ for the mesa suspended in vacuum. (a) $\overline{B}_1/G^{(1)}_{11}=5$ and is $\phi_{01}$ fixed as indicated.  (b) $\overline{B}_1/G^{(1)}_{11}=7$ and $\phi_{01}$ is random.}\label{fig6}
\end{figure}

From the data in Table I and the graph in Fig. 3, it is evident that when the $n=1$ fundamental $ac$ Josephson mode locks onto the cavity $(11)$ mode ($k_J=k'_{11}=k_{11}/n_r$), the cavity $(01)$ mode wave vector $k'_{01}$ only exceeds that of the second $ac$ Josephson harmonic $2k_J$ by $\approx 4$ \%.   Although it will be nearly impossible to excite both the cavity $(11)$ and $(01)$ modes simultaneously, in case the $Q_{01}$ of the cavity $(01)$ is very low, that mode might be very slightly excited, slightly affecting  the output power of the radiation at $2\omega_J$.  The combined electric fields at $2\omega_J=2k_Ja$ after equilibrium has been achieved are given by
\begin{eqnarray}
E_{\theta}&=& [-i\overline{B}_2\sin\theta J_0(2k_{\theta})-\tilde{G}^{(2)}_{01}J_1(2k_{\theta})]\nonumber\\
& &\times\frac{e^{i(k_Jr-\omega_Jt)}}{4\pi r},\label{Etheta2}
\end{eqnarray}
where $\overline{B}_2$ and $\tilde{G}^{(2)}_{01}$ are given by Eqs. (\ref{Bn}) and (\ref{tildeGmp}) in the Appendix.  Note that the off-resonant cavity response to the driving $ac$ Josephson $n$th harmonic differs in phase by $-\pi/2$ from the resonant response, as noted just before Eq. (\ref{tildeGmp}).  Hence, the contributions from the second $ac$ Josephson harmonic current source and the off-resonant $(01)$ cavity mode are out of phase by $\pi/2$.  We note that the combined radiation exhibits $C_{\infty v}$ point group symmetry, rotationally invariant about $\theta=0^{\circ}$, regardless of whether $\phi_{00}$ is fixed or random, and is linearly polarized along  $\hat{\bm\theta}$.   Since both terms vanish at $\theta=0^{\circ}$ and we expect $|\tilde{G}^{(2)}_{01}/\overline{B}_2|\ll1$, the output power is unlikely to differ noticeably from that pictured in Fig. 2(b), except for the substrate factor discussed in Sec. VIII.  We note however, that some azimuthal anisotropy to the second harmonic output power could arise from the spatially inhomogeneous part $\delta J_n^J({\bm x}')$ of the $ac$ Josephson current, as shown in Sec. III.  From Eq. (\ref{EAfull}), the $n=2$ contribution could have various non-vanishing terms proportional to $\cos[m(\phi-\phi_{01})]$, although we expect their amplitudes to be small.

\section{VII. Electric polarization of the combined radiation}

The polarization of the combined output at the fundamental frequency $\omega_J=\omega_{11}$ and $k_J=k_{11}/n_r$, under the assumption that the conditions are appropriate for the fundamental Josephson frequency to excite the $(11)$ cavity mode is more interesting.  After equilibrium has been achieved, the combined electric field in the radiation zone will have the components given in Eqs. (\ref{Etheta1}) and (\ref{Ephi1}).
    The radiation of the combined Josephson $n=1$ fundamental mode locked onto the cavity $(11)$ mode generally has linear  polarization.  This is because the amplitudes of $E_{\theta}$ and $E_{\phi}$ are pure imaginary, so the phase difference between these two components is zero for all times.  However, the axis of polarization depends upon the particular values of $(\theta,\phi)$.  There are some special cases.  These are
 (I)  If $\phi_0$ is random on the time of the measurement, the radiation is linearly polarized along $\hat{\bm\theta}$, arising from the dipole radiation only.  This vanishes at $\theta=0^{\circ}$, of course, so the radiation would be unpolarized there, and one would measure no radiation power through the polarizer. (II)  If $\phi-\phi_{01}=0,\pi$, the polarization is along $\hat{\bm\theta}$. (III) If $\theta=0^{\circ}$, the polarization direction depends strongly upon $\phi-\phi_0$.  But in a measurement at $\theta=0^{\circ}$, one would integrate over all $\phi$ values, measuring an unpolarized radiation, in agreement with experiment\cite{Kadowaki2}. (IV) If $\theta=\pi/2$, the radiation is linearly polarized along $\hat{\bm\theta}$.  (V) At the special line where $E_{\theta}=0$, the polarization is along $\hat{\bm \theta}$.  There is in addition one line given by
 \begin{eqnarray}
 \phi-\phi_{01}&=&\cos^{-1}\Bigl(\frac{\overline{B}_1\sin\theta J_0(k_{\theta})}{G_{11}^{(1)}J_1^{-}(k_{\theta})}\Bigr),
 \end{eqnarray}
 where $E_{\theta}=0$, and the polarization is along $\hat{\bm\phi}$.

 For the general case, the polarization is tilted away from the $\hat{\bm\theta}$ direction by the angle $\tau$ given by\cite{antenna,antenna2}

 \begin{eqnarray}
 \tau&=&\tan^{-1}\Bigl(\frac{E_{\phi}}{E_{\theta}}\Bigr).
 \end{eqnarray}

In Fig. 7, we presented three-dimensional plots of $2\tau/\pi$ in radians versus $\theta$ in degrees and $r=\overline{B}_1/G^{(1)}_{11}$ for $\phi-\phi_{01}=\pi/2$ and $\phi-\phi_{01}=\pi/4$ in the left and right panels, respectively.  In Fig. 7(a), when $\phi-\phi_0{01}=\pi/2$, $E_{\theta}$ is only given by the $ac$ Josephson (or dipole) radiation, which vanishes in the limits $\theta=0^{\circ}$ and $r=0$.  When $\theta=90^{\circ}$, as long as $r>0$, the polarization is along $\hat{\theta}$.  From this figure, it is evident that when $r\approx1$ or greater and $\theta\approx10^{\circ}$ or greater, $\tau\approx0$, and the polarization lies along $\hat{\bm\theta}$.  In  Fig. 7(b), when $\theta=0^{\circ}$ or $r=0$, the polarization is -45$^{\circ}$ between $\hat{\bm\phi}$ and $\hat{\bm\theta}$, which is clear from Eqs. (\ref{Etheta1}) and (\ref{Ephi1}).  Then, as either $r\approx1$ and $\theta\approx10^{\circ}$, the polarization switches to the $\hat{\bm\theta}$ direction, assuming the coherent combination with $\phi_{01}$ is fixed in the measurement.  If $\phi_{01}$ is not fixed, but random, then the  polarization will always be along $\hat{\bm\theta}$, except that one will measure an unpolarized radiation at $\theta=0^{\circ}$.  Preliminary measurements confirmed the unpolarized radiation at $\theta=0^{\circ}$.  However, that single experiment cannot distinguish whether $\phi_{01}$ is fixed or random, because at $\theta=0^{\circ}$, the polarization averages to zero, both for fixed and random $\phi_{01}$.  Hence, further experiments are necessary to distinguish whether $\phi_{01}$ is fixed or random.  A measurement of the degree of coherence could prove useful in this regard.
\begin{figure}
\includegraphics[width=0.235\textwidth]{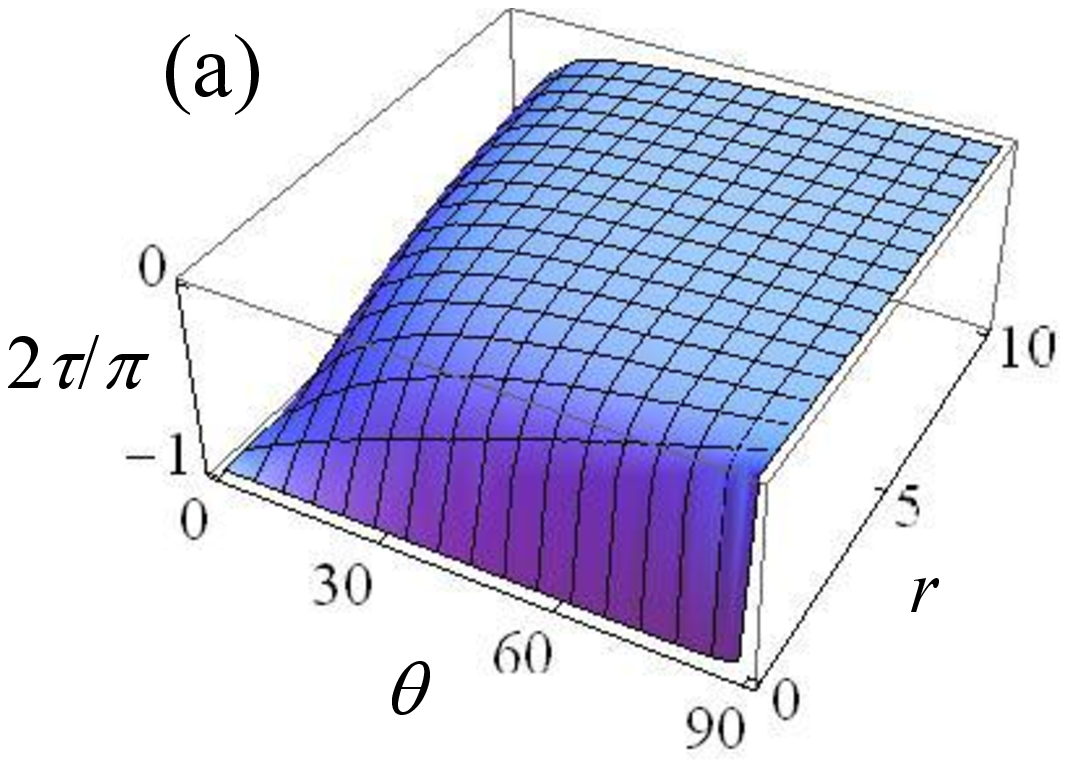}\hskip5pt
\includegraphics[width=0.235\textwidth]{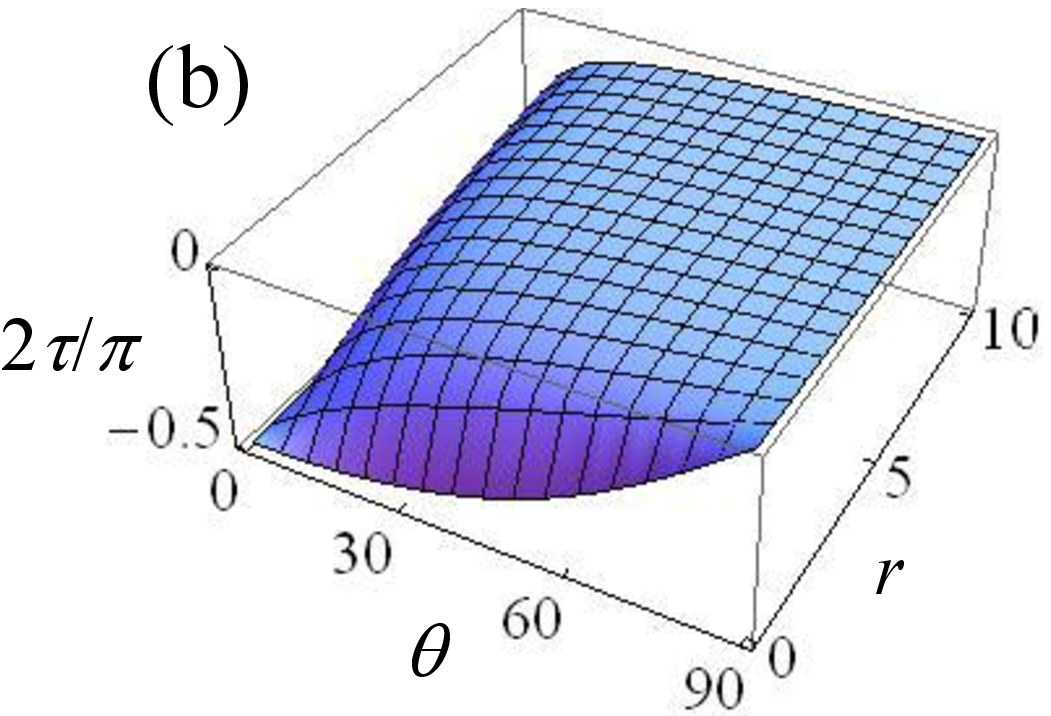}
\caption{(Color online)  Plots of the normalized tilt angle $2\tau/\pi$ in radians that the polarization makes with $\hat{\bm \theta}$ versus $\theta$ in degrees and $r=\overline{B}_1/G_{11}^{(1)}$ of the mixed source radiation for fixed $\phi_{01}$ at the fundamental $n=1$ frequency with $k_{11}a=1.8412/n_r$ when the mesa is either suspended in vacuum or on a superconducting substrate. (a) $\phi-\phi_{01}=\pi/2$. (b) $\phi-\phi_{01}=\pi/4$. .}\label{fig7}
\end{figure}

\hskip-30pt
\begin{figure}\hskip-10pt
\includegraphics[width=0.5\textwidth]{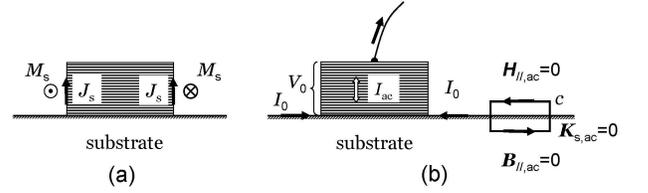}\vskip-20pt
\caption{(a) Sketch of a mesa with  ${\bm J}_S$ and ${\bm M}_S$ surface current sources.  (b) Mesa during coherent emission with applied $dc$ $I_0,$ $V_0$, and $I_{ac}$  confined to it.  Curve $c$ is the integration path for the Amp{\`e}re law boundary condition.  See text.}\label{fig8}
\end{figure}

\begin{figure}
\includegraphics[width=0.25\textwidth]{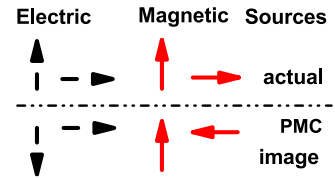}
\caption{(Color online) Sketches of the electric (dashed, black) and magnetic (solid, red) actual and image sources and on a perfect magnetic conductor substrate. From Ref. (2).}\label{fig9}
\end{figure}

\section{VIII. Superconducting substrates}

We now discuss the consequences of mixed sources of the radiation.  If a cavity mode were the only source of the radiation, the output power would vanish at $\theta=90^{\circ}$.  This is in close agreement with experiments on rectangular and cylindrical mesas\cite{Kadowaki2,Kadowaki3}.  However, if the fundamental $ac$ Josephson current is also a radiation source, it would have a contribution at $\theta=90^{\circ}$ that would be non-vanishing.  Since our qualitative fits to the data on both types of mesas suggest that the majority of the radiation source arises from the electric surface current (or the $ac$ Josephson current), it then is important to discuss why the experiments see a maximum output at $\theta\approx30^{\circ}$, but a seemingly vanishing output power at $\theta=90^{\circ}$.

Here we argue that the $ac$ Meissner effect in the superconducting substrate sketched in Fig. 8 (a) causes it to behave as a perfect magnetic conductor (PMC)\cite{antenna,antenna2,KK}, which can be treated by the image source technique as sketched in Fig. 9. Note that in our treatment of the Love equivalence principle, we assumed that the magnetic conductor model allowed us to represent the magnetic field internal to the mesa as a surface electric current, and to set ${\bm H}=0$ on the surface of the mesa.  Present mesa antennas atop a PMC substrate emit  dramatically lower output power than those atop a substrate that is an insulator or perfect electric conductor (PEC)\cite{antenna,antenna2}.  We note that a superconducting Nb substrate was previously used as a ground plane for a square lattice of Nb/Al$_2$O$_3$ Josephson junctions, the output of which was amplified by a cavity at the edge of the array\cite{Barbara1,Barbara2,Vasilic}.  Without the ground plane, the radiation was not detectable.  In that case, the $ac$ Josephson current source was parallel to the substrate \cite{Barbara1,Barbara2,Vasilic}, so that the image of the supercurrent in the magnetic conducting substrate was parallel to the source current, as sketched by the two arrows second from the left in Fig. 9.  In our case, both the electric and magnetic image currents are opposite to the respective source currents, as in the far left and far right arrow pairs in Fig. 9, causing a cancellation of the output power at $\theta=90^{\circ}$.  For pure cavity mode radiation at the precise mode frequencies, the output would vanish there anyway, but for the electric current ($ac$ Josephson) source, it would not.  A study of the angular dependence of the second harmonic would therefore be a critical test of this substrate model.

This unexpected combination of $I(90^{\circ})\approx0$ and $\theta_{\rm max}\approx30^{\circ}$ led us to consider the substrates of existing samples\cite{Ozyuzer,Kadowaki}.  Each existing mesa is a structure formed by ion milling on the top center  of a wider and thicker single crystal of superconducting BSCCO\cite{Ozyuzer}.  We neglect the thin top Au mesa contact.  As sketched in Fig. 8 (b), during coherent Josephson radiation, the $ac$ Josephson current is essentially confined to the mesa by the applied $dc$ current $I_{0}$ and voltage $V_{0}$.   With only a $dc$ surface current density $\propto I_0$ and  ${B}_{||,ac}(t)=0$ beyond the skin depth ($\approx0.15$~$\mu$m ) inside the BSCCO substrate due to the $ac$ Meissner effect, the Amp{\`e}re-Maxwell boundary condition forces  ${H}_{||,ac}(t)=0$ just above the BSCCO substrate\cite{Jackson}.  This corresponds to a perfect magnetic conductor (PMC) substrate, with the effective image sources sketched in Fig. 5\cite{antenna,antenna2}. Thus, for a BSCCO substrate, we restrict $\theta$ to $0\le\theta\le90^{\circ}$ and replace $\eta(z')$  in Eqs. (\ref{MS}) and (\ref{JS}) by
 \begin{eqnarray}
 \eta_{-}(z')&=&{\rm sgn}(z')\Theta[h^2-(z')^2].\label{substrate}
 \end{eqnarray}
   For cylindrical  mesas in the radiation zone,  $h<<a,r$, and we assume $h\ll 1/k_n$ for the relevant $n$.  One may expand $e^{ik_nR}/R$ in the generalization of Eq. (\ref{F}) for small $z'$.   From Eq. (\ref{substrate}), the electric and magnetic current substrate factors are simply obtained from terms linear in $z'$.  In the radiation zone,
\begin{eqnarray}
S^J_n(\theta)&{{\rightarrow}\atop{r\rightarrow\infty}}&-ik_nh\cos\theta\>\>\Theta(90^{\circ}-\theta),
\end{eqnarray}
and $S^M_{mp}(\theta)$ is obtained from $S^J_n(\theta)$ by $k_n\rightarrow k_{mp}$, where $(mp)$ applies either to the cavity mode, which can be either cylindrical or rectangular, as disccussed in Sec. IX.  For the case of the tail of a broad cavity $(mp)$ mode overlapping the  $n$th $ac$ Josephson harmonic, we replace $S^M_{mp}(\theta)$ with $S^M_n(\theta)$, as in Eq. (\ref{tildeGmp}) in the Appendix.

\begin{figure}
\includegraphics[width=0.235\textwidth]{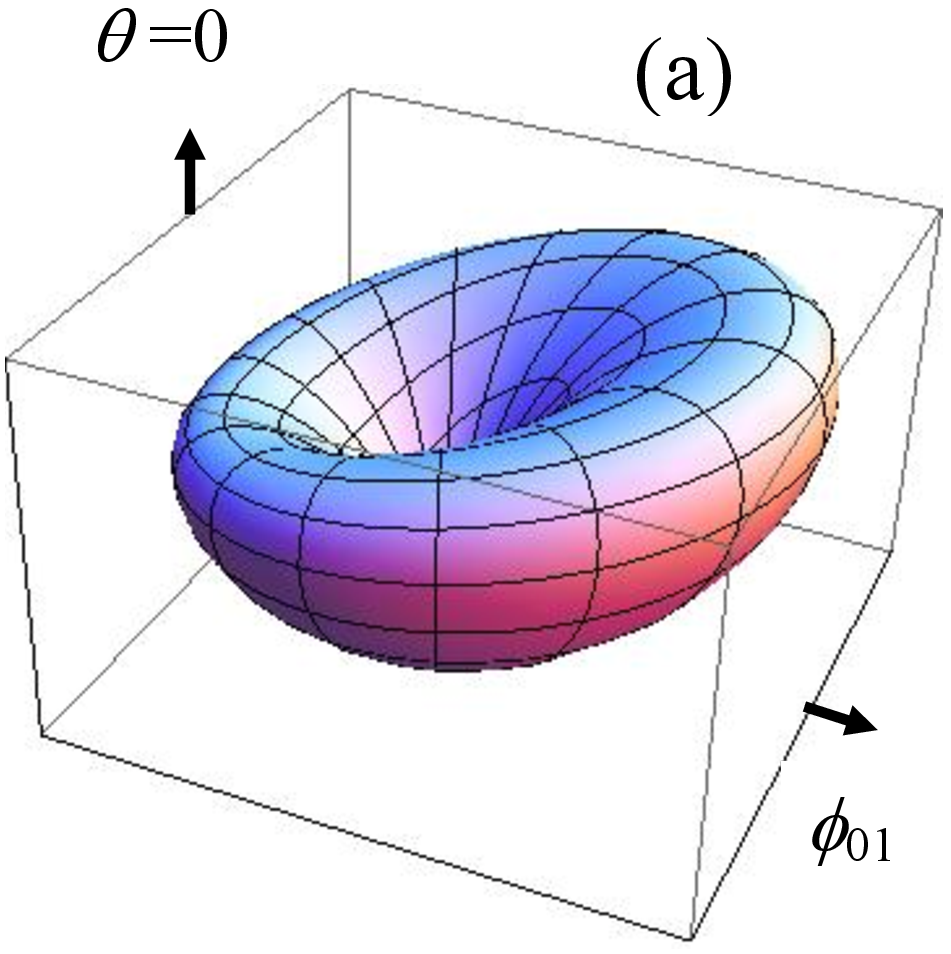}\hskip5pt
\includegraphics[width=0.235\textwidth]{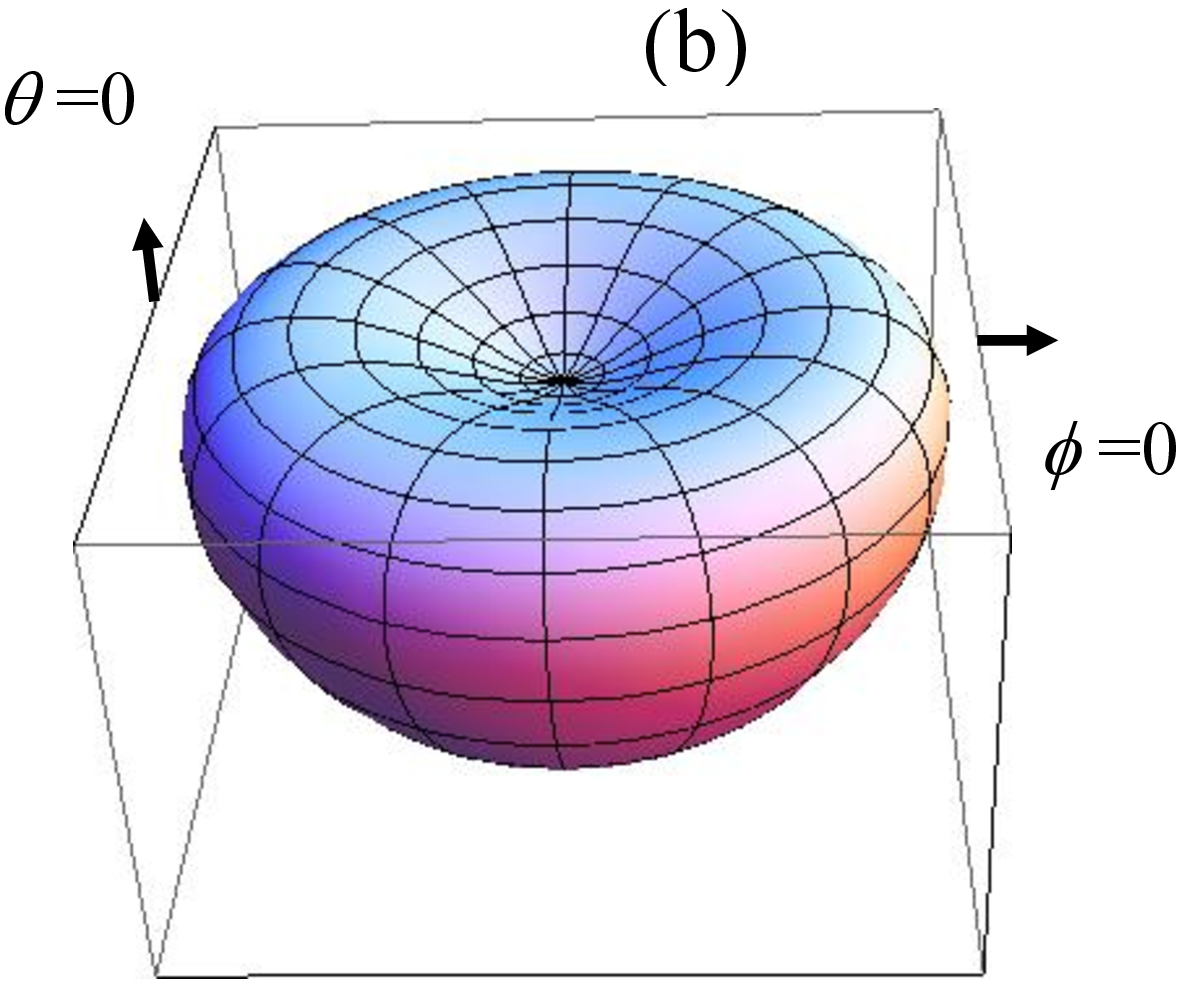}
\caption{(Color online) Plots of  the  radiation power emitted from the mixed $n=1$ $ac$ Josephson and the $(11)$ cylindrical cavity  modes with  $k_{11}a=1.8412/n_r$  fixed and the mesas sit atop a superconducting substrate. (a) $\phi_{01}$ is fixed as indicated, and $\overline{B}_1/G_{11}^{(1)}=7$.   (b) $\phi_{01}$ is random and $\overline{B}_1/G^{(1)}_{11}=1$.}\label{fig10}
\end{figure}

\begin{figure}
\includegraphics[width=0.234\textwidth]{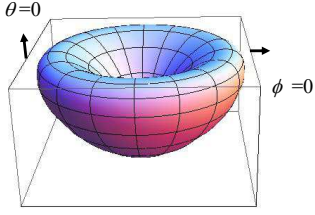}
\caption{(Color online) Plot of the radiation power at the $n=2$ second harmonic of the $ac$ Josephson current for a cylindrical mesa with $k_2a=2*1.8412/n_r$ on a superconducting substrate.}\label{fig11}
\end{figure}

In  Figs. 10 and 11, we show our predictions for  the conducting dipole model with $n=1, 2$ respectively, of a cylindrical mesa atop a superconducting substrate.   Note that in each case, the radiation vanishes at $\theta=90^{\circ}$, unlike any of the radiation from the same modes shown in the left panels when the mesa is suspended in vacuum.  The substrate factor makes a dramatic effect for $\theta\approx90^{\circ}$, causing it to vanish  at $\theta=90^{\circ}$, in apparent agreement with experiment.  In Fig. 10(a), we assumed $\phi_{01}$ was fixed, and $\overline{B}_1/G_{11}^{(1)}=7$, which maintains the point kink found for the combined output pictured in Fig. 6(a) for the mesa suspended in vacuum.  In Fig. 10(a), the predicted radiation vanishes at $\theta=90^{\circ}$, but it is still highly anisotropic for $\theta<90^{\circ}$, with a point kink that is off center (not at $\theta=0^{\circ}$).  Experiments so far cannot eliminate this possibility, because scans across the diameter of the cylinder have only been made on two perpendicular planes, and it is possible that such a point kink would have been missed. In Fig. 10(b), $\overline{B}_1/G_{11}^{(1)}=1$, $\phi_{01}$ is random, the point kink is removed, and the output power has a maximum at $\theta\approx 34.5^{\circ}$.  These predictions are remarkably similar to the experimental results\cite{Kadowaki3}. In Fig. 11, we showed the modification to Fig. 2(b), the second $ac$ Josephson harmonic at $n=2$, due to the superconducting substrate.  Since the $(01)$ cavity mode is significantly far removed from this $ac$ Josephson mode, this should represent the output power a the second harmonic.  It has a maximum at $\theta\approx42.2^{\circ}$.

\section{IX. Rectangular mesas}
Finally, we come back to the more complicated rectangular mesas.  A sketch of the electric and magnetic surface current densities for the $(10)$ rectangular mesa cavity mode is shown in Fig. 12.  Although the surface electric current density ${\bm J}_S$ is uniform along the edge of the mesa, the magnetic surface electric current changes its direction by $90^{\circ}$ at each of the corners.  Hence, a precise analytic treatment of $\delta J_n({\bm x}')$ within the rectangular mesa, subject to the correct boundary condition that the tangential component ${H}_{||}$ of the magnetic field vanishes on each of the mesa edges, is complicated by the presence of the corners.  However, we shall ignore such complications, and focus of the radiation that arises when ${\bm J}_S$ and ${\bm M}_S$ are placed on the mesa edges.

\begin{figure}
\includegraphics[width=0.45\textwidth]{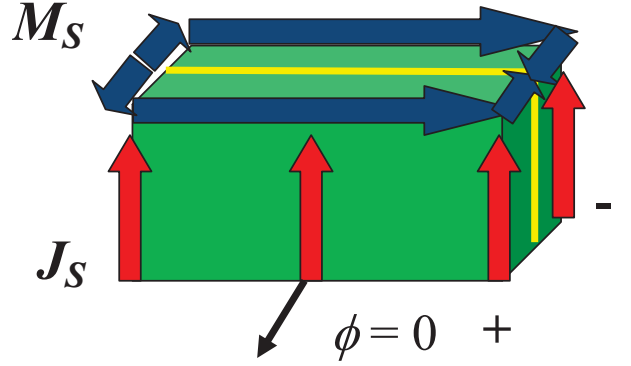}
\caption{(Color online) Sketch of a rectangular mesa with a surface electric current density ${\bm J}_S$ (red vertical arrows) and a magnetic current density ${\bm M}_S$ (blue horizontal azimuthal arrows).  The $\pm$ signs refer to the signs of the rectangular $(10)$ cavity mode, separated by the yellow line.}
\end{figure}\label{fig12}

We now consider a rectangular mesa of width $w$, length ${\ell}$, and height $h$ in vacuum satisfying ${\ell}>\lambda_c>w$, where $\lambda_c \sim 170$~$\mu$m is the $c$-axis penetration depth.  In this case, the rectangular mesas allow Josephson vortices parallel to the mesa widths, suggesting that the only stable cavity modes that could be excited are the  TM$^z_{n00}$ modes, which are constant in position along the mesa height and length, and oscillate in position with integral multiples of half-wavelengths along the mesa widths. In rectangular source coordinates $(x',y',z')$,
\begin{eqnarray}
{\bm J}_S({\bm x}')&=&\hat{\bm z}'\frac{J_J}{4}\eta_J(z')\sum_{\sigma=\pm}[f_{\sigma}(x',y')+g_{\sigma}(x',y')],\label{Jrect}\\
{\bm M}_{Sn}({\bm x}')&=&\frac{\tilde{E}_{0n}}{4}\eta_M(z')\sin[n(x'-x_n)\pi/w]\nonumber\\
& &\times\sum_{\sigma=\pm}\sigma[\hat{\bm y}'f_{\sigma}(x',y')-\hat{\bm x}'g_{\sigma}(x',y')],\label{Mrect}\\
f_{\sigma}(x',y')&=&w\delta(x'+\sigma w/2)\Theta[(\ell/2)^2-(y')^2],\label{f}\\
g_{\sigma}(x',y')&=&{\ell}\delta(y'+\sigma \ell/2)\Theta[(w/2)^2-(x')^2],\label{f}
\end{eqnarray}
where the TM$^z_{n00}$ cavity mode energy is degenerate for $-w/n\le x_n\le w/n$.

 Since standard cavity mode boundary conditions may be altered when the corners are treated correctly, we  treat the output power obtained from the combination of the ${\bm J}_S$ and ${\bm M}_{Sn}$ sources in three models.  In the first model, we treat the combination as coherent, satisfying $H_y(x'=\pm w/2)=0$.  This leads to the solutions $x_n=0, w/n$ for $n$ odd, and $x_n=\pm w/2n$ for $n$ even. These cases lead to an asymmetry and kinks in the angular distribution of the output power. In the other two models, the output from the two sources is incoherent.  In Model I, we average the output power $P(x_n)$  over the two waveforms with the same boundary conditions, $\langle P(x_n)\rangle_{I}=\frac{1}{2}[P(0)+P(w/n)]$ for $n$ odd, and $\langle P(x_n)\rangle_{I}=\frac{1}{2}[P(w/2n)+P(-w/2n)]$ for $n$ even\cite{antenna}. This model assumes the standard cavity model applies, but for mixed sources, we assume that both possible phases (such as the signs in Fig. 12) of the electric field are allowed with equal probability.  This is the rectangular cavity analogue of the random $\phi_{0n}$ model for the emission from a cylindrical cavity.  In Model II, we let $\langle P(x_n)\rangle_{II}=\int_{-w/n}^{w/n}dx_nP(x_n)n/2w$.  This model relaxes the standard cavity prescription, but preserves the wavelengths across the mesa.

 Then, the time-averaged power per unit solid angle in the radiation zone is
 \begin{eqnarray}
 \frac{dP}{d\Omega}&{{\rightarrow}\atop{r/a\rightarrow\infty}}&\frac{Z_0(J_J\tilde{v}k_1)^2}{128\pi^2}\sum_{n=1}^{\infty}n^2\biggl[\Bigl|\sin\theta a_n\chi_n(\theta,\phi)S^J_n(\theta)\Bigr|^2\nonumber\\
 & &\>+\alpha_n(\theta)\Bigl(C^i_n+D^i_n-\sin^2\theta[C^i_n\cos^2\phi\nonumber\\
 & &\qquad+D^i_n\sin^2\phi-E^i_n\sin\phi\cos\phi]\Bigr)\biggr],\label{Prect}
 \end{eqnarray}
 where $i=$ I, II, $\alpha_n(\theta)=|\tilde{E}_{0n}S^M_n(\theta)|^2/(2Z_0J_J)^2$, and $\chi_n(\theta,\phi)$ and the $C_n^i(\theta,\phi)$, $D_n^i(\theta,\phi)$, and $E_n^i(\theta,\phi)$ are given in the Appendix.  Our results for the angular dependence of the output power from rectangular mesas is shown in  Figs. 13-15.  These figures are for rectangular mesas with the length normal to $\phi=0$.   Although the figure ``boxes'' appear to have the short length along $\phi=0$, that is the direction in which the radiation is largest.  Since these figures correspond to the $(m0)$ modes for $m=1,2$, the radiation is primarily along the length of the mesa, along or near to $\phi=0$.  In the figures that follow, we assume ${\ell}/w=20/3$, as in some of the mesas\cite{Ozyuzer,Kadowaki}.  Three-dimensional plots of $I(\theta,\phi)\propto dP(\theta,\phi)/d\Omega$ in arbitrary units are then obtained.  First, we present the data for the emission from the primary source, the $ac$ Josephson current in the form of the surface electric current density ${\bm J}_S$.  In Fig. 13, we present $I(\theta,\phi)$ for the $n=1$ and $n=2$ $ac$ Josephson radiation, respectively for the mesa suspended in vacuum.   The corresponding predictions for the rectangular cavity $(10)$ and $(20)$ modes at the same frequencies as the $n=1,2$ $ac$ Josephson radiation are shown for Model I in Fig. 14 for the cavity suspended in vacuum. We note that the output for the rectangular cavity $(10)$ mode pictured in Fig. 14(a) is very similar to that for the cylindrical cavity $(11)$ pictured in Fig. 4(a) when both mesas are suspended in vacuum. We note that the angular dependence of the second harmonic emitted from rectangular cavities should be very interesting to measure, especially as it contains a mixture of the radiation predicted in Figs. 13(b) and 14(b), and is distinctly different in form from any of the cylindrical cavity modes calculated without a substrate.

  \begin{figure}
\includegraphics[width=0.235\textwidth]{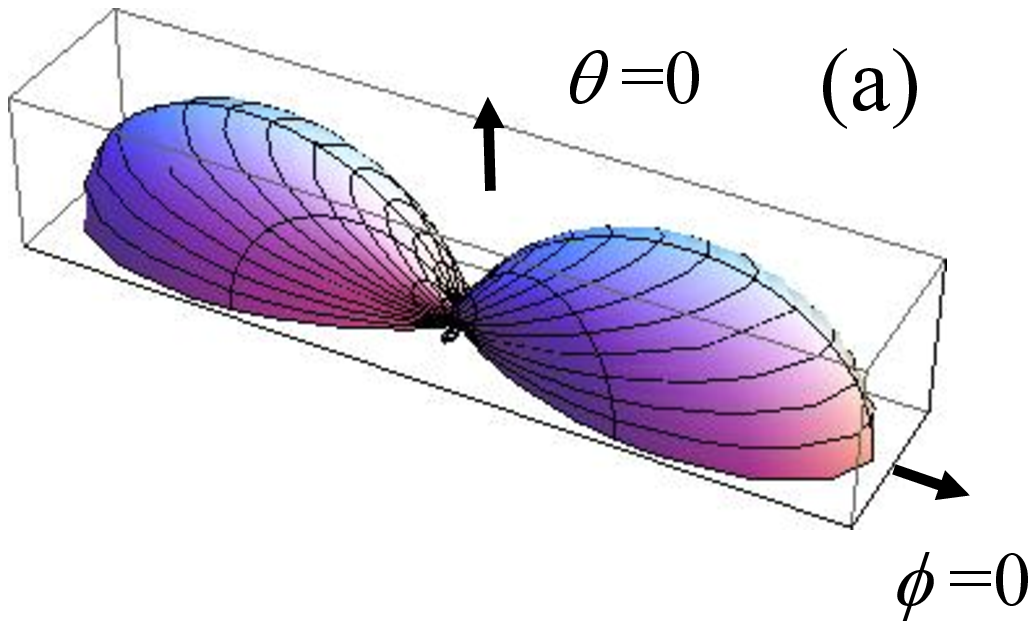}\hskip5pt
\includegraphics[width=0.235\textwidth]{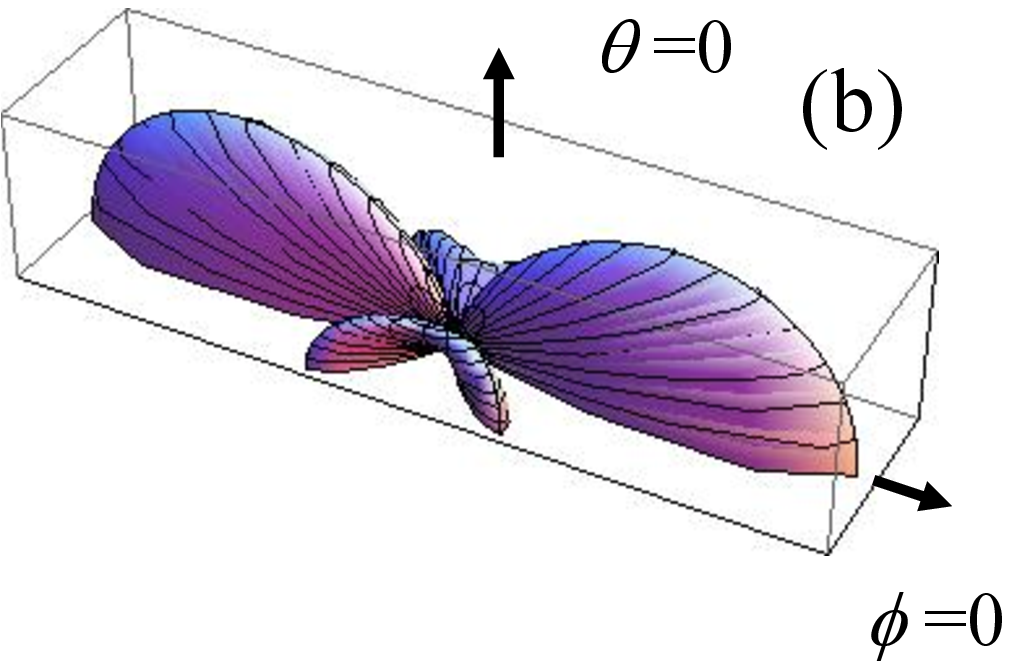}
\caption{(Color online)  Three-dimensional plots of  the radiation intensity  in arbitrary units for rectangular mesas with ${\ell}/w=20/3$ from the uniform $ac$ Josephson current alone, when the mesa is suspended in vacuum.  (a) At the fundamental $n=1$ frequency with $k_1w=\pi/n_r$. (b) At the second harmonic $n=2$ with $k_2w=2\pi/n_r$.}\label{fig13}
\end{figure}
 \begin{figure}
\includegraphics[width=0.235\textwidth]{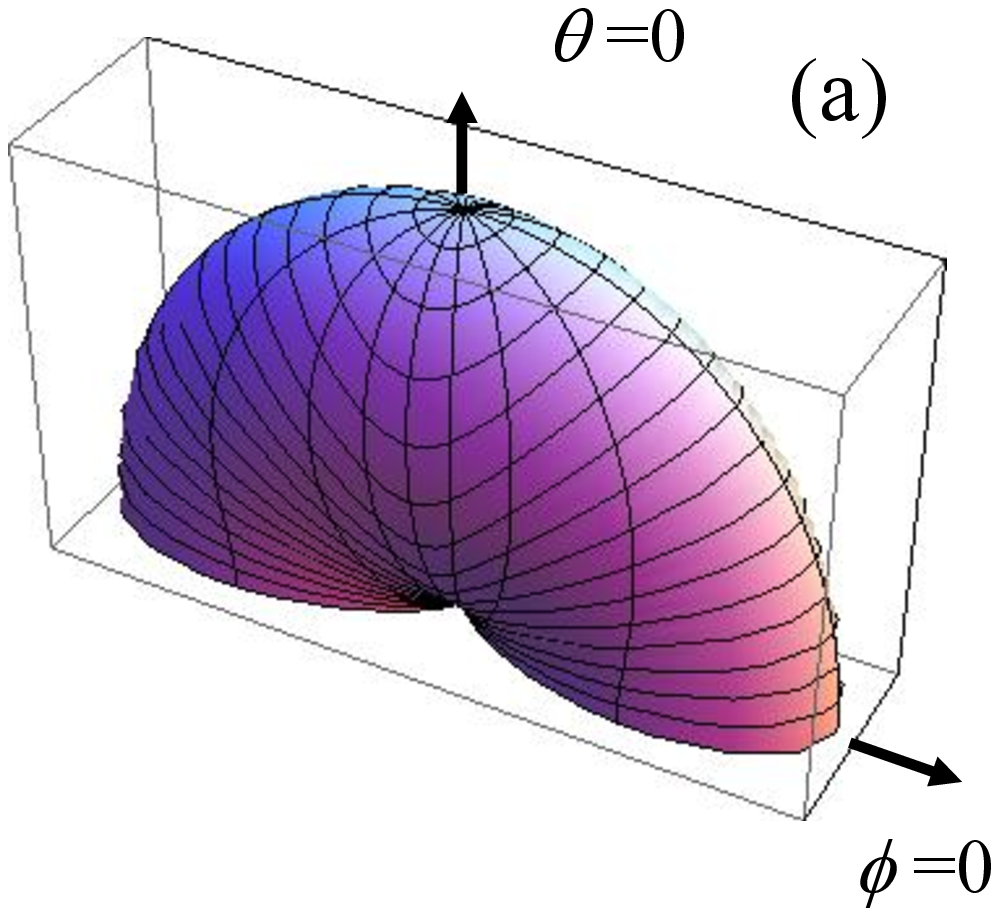}\hskip5pt
\includegraphics[width=0.235\textwidth]{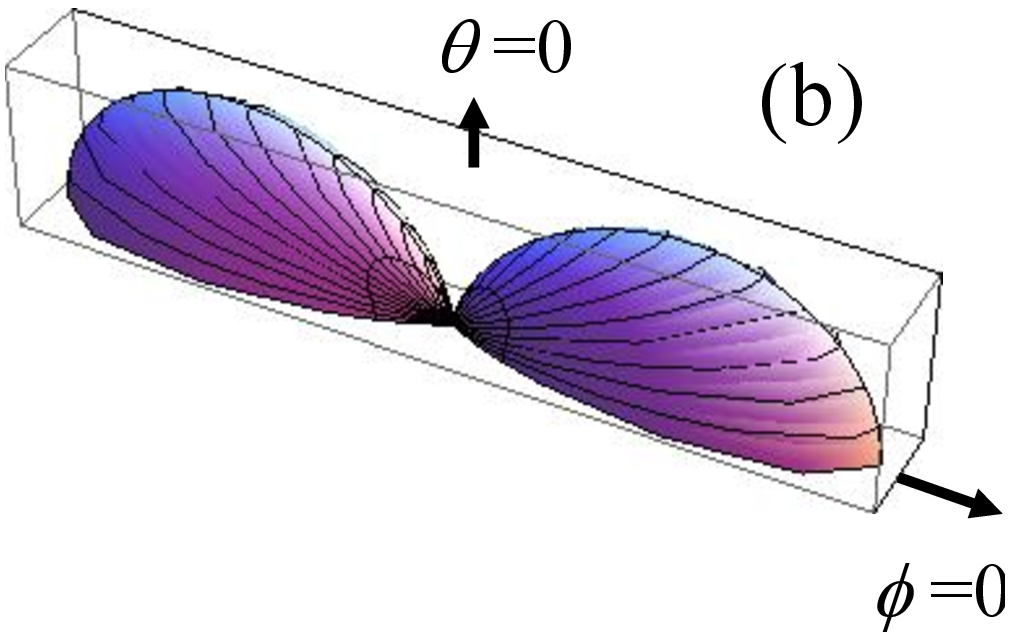}
\caption{(Color online)  Three-dimensional plots of  the radiation intensity in arbitrary units for rectangular mesas with ${\ell}/w=20/3$ from the cavity source alone, when the mesa is suspended in vacuum. (a) The fundamental $(10)$ cavity mode with $k_{10}w=\pi/n_r$. (b) The second harmonic $(20)$ cavity mode with $k_{20}w=2\pi/n_r$.}\label{fig14}
\end{figure}

 \begin{figure}
\includegraphics[width=0.235\textwidth]{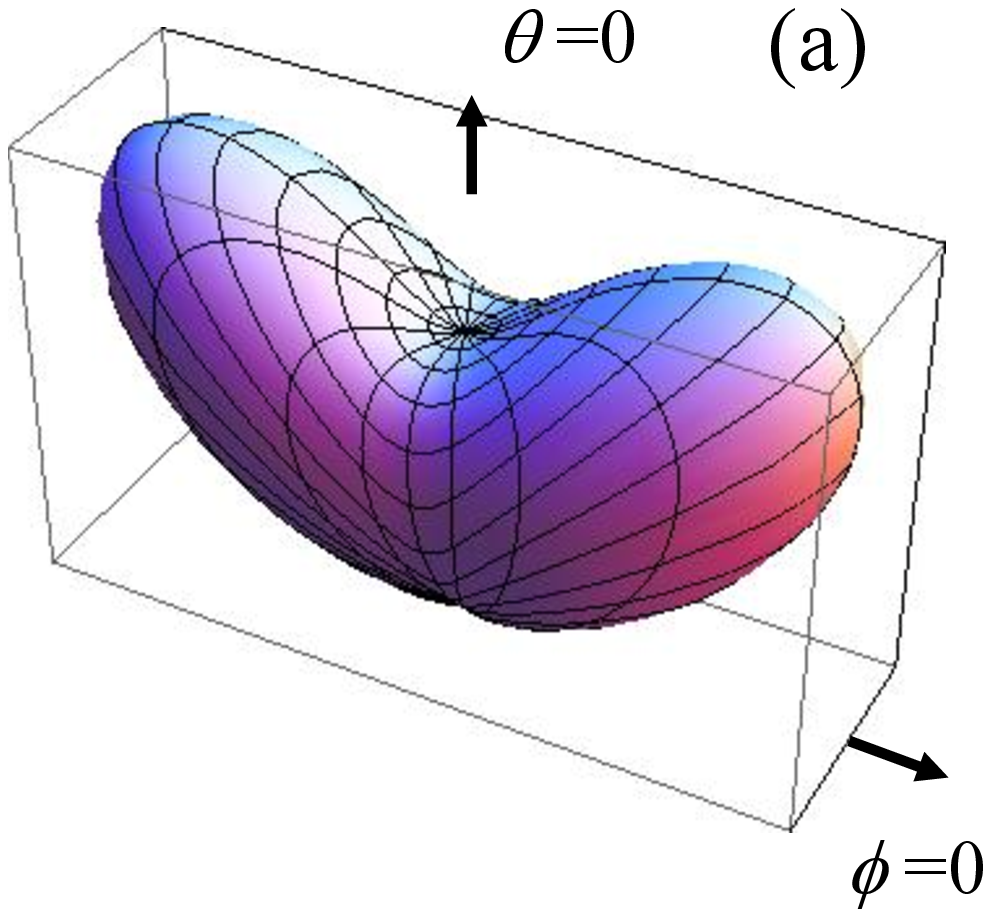}\hskip5pt
\includegraphics[width=0.235\textwidth]{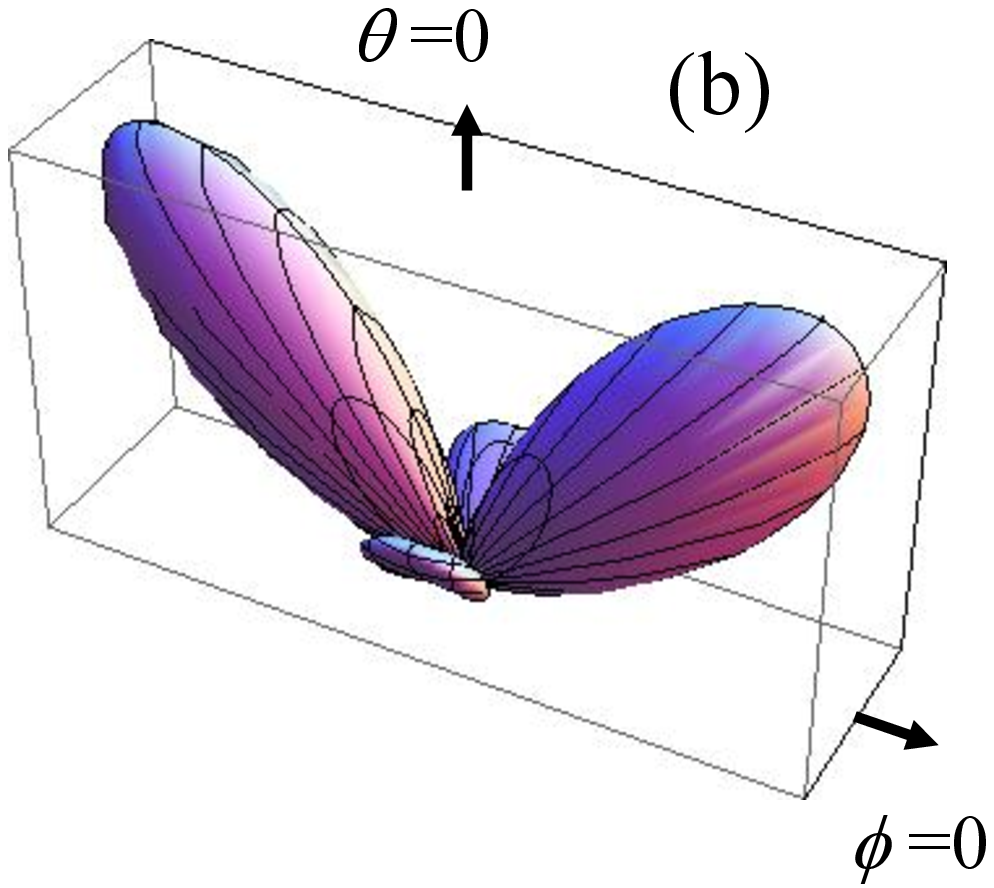}
\caption{(Color online)  Plots of  the  intensity  in arbitrary units for rectangular mesas with ${\ell}/w=20/3$ of the combined radiation from Model I with $\alpha(0)=0.2$ from the uniform $ac$ Josephson current and the rectangular cavity modes when the mesa sits atop a superconducting substrate. (a) The fundamental $n=1$ $ac$ Josephson mode and the cavity $(10)$ mode with $k_w=k_{10}w=\pi/n_r$.  (b) The second harmonic with $k_2w=k_{20}w=2\pi/n_r$. }\label{fig15}
\end{figure}

In Fig. 15, we show the combined output from the primary $ac$ Josephson current and excited cavity modes for a rectangular mesa atop a superconducting substrate.  In Fig. 15(a), the fundamental $n=1$ mode is locked onto the rectangular cavity $(10)$ mode, and the figure assumes $\alpha(0)=0.2$ using Model I for the cavity.   This figure  is very similar to that observed experimentally\cite{Kadowaki2}. In Fig. 15(b), the predicted output power is shown for the second harmonic, again using Model I with $\alpha(0)=0.2$ for a mesa atop a superconducting substrate.

As for cylindrical mesas, the output from rectangular mesas is linearly polarized.  Because the angular distribution of the output intensity at the fundamental frequency is very similar in both rectangular and cylindrical mesas, only minor differences in the angular distribution of the tilt angles for the two types of mesas are expected.

 \begin{figure}
\includegraphics[width=0.235\textwidth]{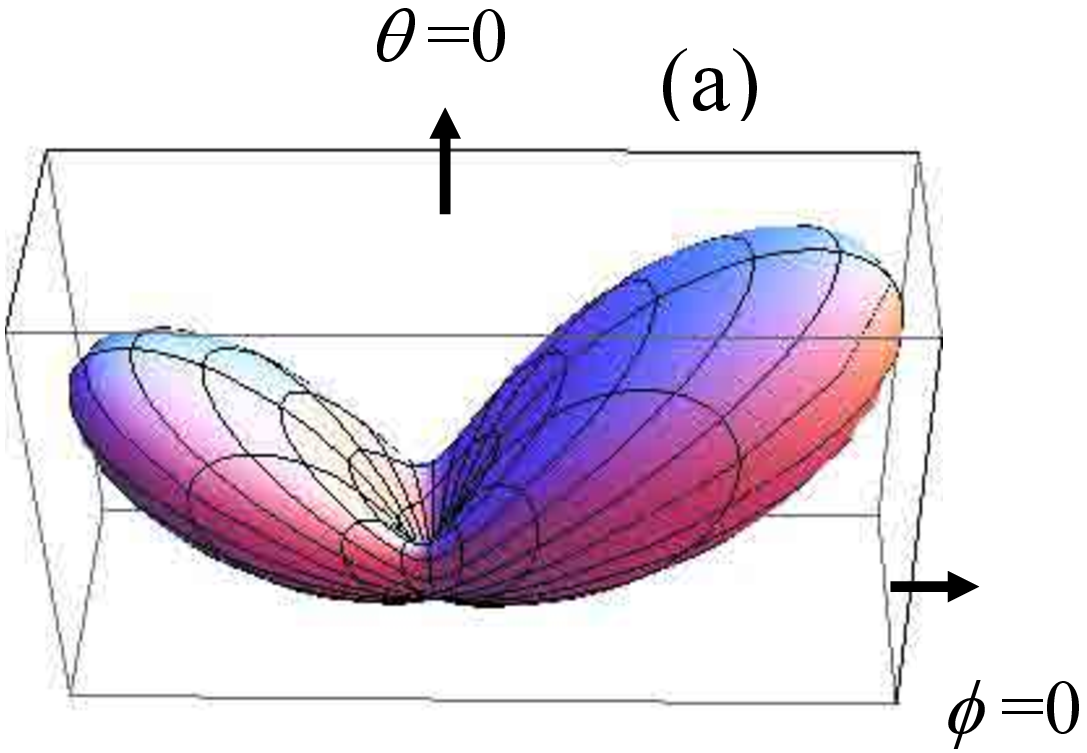}\hskip5pt
\includegraphics[width=0.235\textwidth]{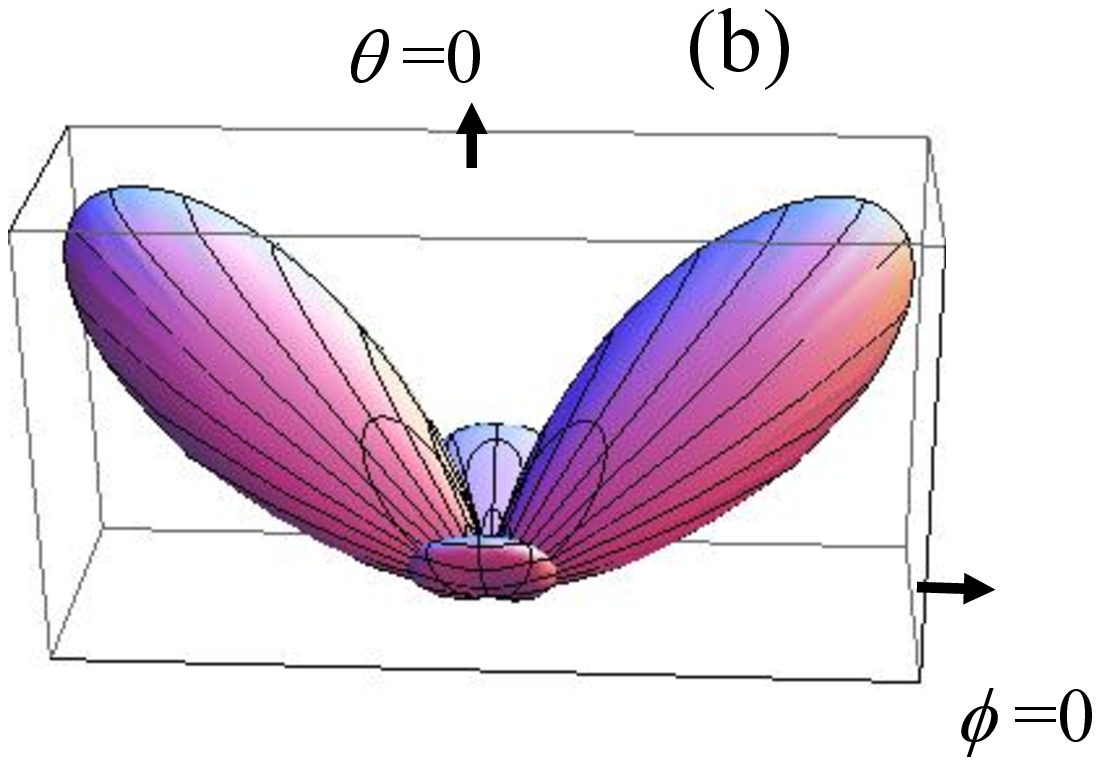}
\caption{(Color online)  Plots of  the  intensity  in arbitrary units for rectangular mesas with ${\ell}/w=20/3$ of the combined radiation  when the mesa sits atop a superconducting substrate. (a) The fundamental $n=1$ $ac$ Josephson mode and the cavity $(10)$ mode with $k_w=k_{10}w=\pi/n_r$, $x_1=0$ and $r_1=0.05$  (b) The second harmonic with $k_2w=k_{20}w=2\pi/n_r$, $x_2=w/4$ and $r_2=0.6$. }\label{fig16}
\end{figure}

We now consider the case when the combination of the radiation sources is coherent.  In this case, the output power is given by
\begin{eqnarray}
\frac{dP}{d\Omega}&\propto&|-\sin\theta\chi_n+r_n(M_n^x\sin\phi-M_n^y\cos\phi)|^2\nonumber\\
& &+r_n^2|M_n^x\cos\phi+M_n^y\sin\phi|^2,
\end{eqnarray}
where $r_n=\tilde{E}_{0n}S_n^M(\theta)/[2\mu_0J_Ja_ncS_n^M(\theta)]$ is real.  For $n=1$, $M_1^x$ and $M_1^y$, given in the Appendix, when evaluated at either $x_1=0$ or $w$ are real, and a coherent combination is strongly asymmetric.  For $n=2$, however, both $M_2^x$ and $M_2^y$ are pure imaginary for  $x_2=\pm w/4$.  Hence, the resulting coherent combination of the output power is a sum of the two output powers from the two sources.  This is pictured in Fig. 16.
 In Fig. 16(a), we show the combined coherent output when $r_1=0.05$, computed when $x_1=0$. This figure shows a strong intensity asymmetry even for this small ratio of the outputs from the two sources, with the pattern exhibiting $C_1$ symmetry.  In Fig. 16(b), the combined output for $n=2$ is shown for the case $r_2=0.6$,  computed with $x_2=w/4$. In this case, the pattern has $C_{2v}$ symmetry.

\section{X. Discussion and Summary}

In this paper, we have presented the first correct prescription of how to incorporate the Amp{\`e}re boundary condition into the superconducting stack of Josephson junctions under the application of a $dc$ voltage $V$, which itself generates THz radiation.  We have used both of the Love equivalence principles. By the Love equivalence principle for a magnetic conductor, the radiation from the magnetic field inside the mesa generated by the $ac$ Josephson current is effectively set equal to zero inside the cavity, and replaced by the equivalent surface electric current ${\bm J}_S$, which is the primary radiation source.   We remark that setting ${\bm H}=0$ inside the mesa also sets the boundary condition for the cylindrical cavity.  With this boundary condition, the cylindrical cavity modes are immediately found, and when the $ac$ Josephson current effectively placed on the edge as ${\bm J}_S$ radiates at a frequency that matches one of the cavity modes, the inhomogeneous part of the $ac$ Josephson current has a mode which resonates with a cavity mode of the same (or perhaps similar, in an extension of our model) spatial form, and the amplitude for that cavity mode then grows linearly in time, until it saturates, and then radiates in conjunction with the primary $ac$ Josephson radiation at that frequency.

The form of the cavity radiation is obtained from Love's electric conductor equivalence principle, in which the equivalent of the cavity electric field is placed on the surface as a magnetic surface current density ${\bm M}_S$, which radiates in conjunction with the $ac$ Josephson radiation in ${\bm J}_S$, the inhomogeneous part of which excited the cavity resonance.  From the experiments on three cylindrical mesas, we conclude that the lowest energy cavity $(11)$ mode has been excited.  In addition, the harmonic radiation at twice and three times the fundamental frequency cannot be explained from the cavity alone, as the mismatch in frequencies is 4\% for the second harmonic, and about 3\% for the third harmonic.  This harmonic radiation therefore arises almost entirely from the primary radiation source, the $ac$ Josephson radiation acting as a surface electric current density.  In addition, the fact that the superconductor acts both as an electric conductor and as a magnetic conductor provides a mechanism to understand the role of a superconducting BSCCO substrate.  Such substrates cause the emitted radiation to vanish along the direction parallel to it, greatly reducing the output power of the radiation.

We remark that while the main function of the inhomogeneous $ac$ Josephson current is that it couples to a cavity mode, provided that the conditions are properly met, if the inhomogeneities are sufficiently strong, they can significantly alter the azimuthal and axial anisotropy of the combined radiation at the fundamental frequency, and especially that of the higher harmonics.  In addition, it remains an open question as to whether the angular constants $\phi_{0n}$ are fixed or random during the time of the measurement.  To the extent that the cylindrical cavity itself is perfectly homogeneous, one might expect no preferred angle for each mode of inhomogeneity.  On the other hand, some feature involved in the experimental situation might lead to a fixed inhomogeneity direction.  Only experiments measuring the angular distribution of the radiation and its polarization can distinguish these two situations.  Although we did not show any figures to illustrate this point, if one were to assume the phase of the cavity mode for a fixed $\phi_{0n}$ could have either sign with equal probability, then the situation would be analogous to our Model I for a rectangular mesa. For a cylindrical cavity, this model would preserve the point kink of the output radiation and restore $C_{2v}$ point group symmetry, but the position of the kink would be at $\theta=0^{\circ}$, and the value of the output power would not vanish there.

For rectangular mesas, setting $\hat{\bm n}\times{\bm H}=0$ on each of the edges leads to singularities in the analytic properties of the mathematical difficulties at the corners, greatly complicating the situation.  Hence, most previous workers studied infinite strips, where the corners could be neglected.  For the case of finite length rectangular mesas, the question of whether the combined radiation is skewed to one side of the mesa or the other is also relevant.  If the mesa were symmetric, as in a cylinder with no preferred angle, one would average over the two configurations, resulting in a symmetric radiation pattern, and an incoherent combination of the radiation from the two sources, each of which is separately coherent.  That was the purpose of our study of Models I and II.

We remark that we have treated the cylindrical mesa essentially as an infinitely thin disk, except in calculating the superconducting substrate factor, completely neglecting any spatial variation normal to the layers.  While this is not a good approximation for treating heating inhomogeneities, it suffices for impurity inhomogeneities and the important lateral inhomogeneities in the $ac$ Josephson current, which we assume to be the critical part of the microscopic mechanism that excites the cavity modes.  Hence, exotic features such as inhomogeneous kinks along the $z$ axis are omitted entirely, as they are completely unnecessary for the excitation of the cavity modes, but instead complicate the situation unnecessarily\cite{note}.

We emphasize that  cylindrical mesas are much simpler to understand theoretically than are rectangular mesas.  There are several aspects to this simplicity.  Most important, the von Neumann boundary condition is trivial in this geometry, whereas for a rectangle, it is practically intractable at the corners, except by numerical techniques.  In addition, the fact that none of the cylindrical cavity modes are harmonics of one another is an extremely important point, which is why we set out to study cylindrical mesas in the first place.  Since rectangular cavities have higher energy modes that are also harmonics of a lower mode, such as the $(10)$ or the $(01)$ mode, the observation of harmonics in the output of rectangular mesas did not allow for a precise determination of the primary radiation source, which has now been clearly identified as the $ac$ Josephson current, and not the cavity mode radiation. The primary role of the cavity mode is to lock an $ac$ Josephson mode (usually, if not always, the fundamental) onto a well-defined cavity mode, fixing $\omega_J$ to $\omega_{m_0p_0}$.  Then, the laterally inhomogeneous $ac$ Josephson current excites the $(m_0p_0)$ cavity mode, allowing it also to radiate.  From fits to experimental data on both rectangular and cylindrical cavities, it appears that at least half of the intensity at the angles $\theta_{\rm max}\approx30-40^{\circ}$ of maximum output arises from the primary source.   This primary radiation source radiates at all of the $ac$ Josephson harmonic frequencies, and to date, the second harmonic has been observed in emission from all three cylindrical mesas under study, and the third harmonic was visible in two of the three mesas studied.  The amplitudes relative to that of the fundamental of the higher harmonics are comparable to those in rectangular mesas, for which amplification by the excitation of higher cavity modes could occur.

 The cavity modes of a cylindrical mesa are similar in form to those of a drum, and one or more of them can be amplified when a drummer strikes a particular spot on the drum surface.    The striking surface region of the drum does not have to have the precise shape of that of the main cylinder (drum)  mode excited.  By analogy, a modification of our theory of the cylindrical cavity mode amplification could occur without implementation of the von Neumann boundary condition to provide a precise matching in spatial form  of the cavity mode with that of the inhomogeneities.  In a rectangular mesa, a subset of the cavity mode frequencies are harmonics of one another, and can be excited as by a player of a stringed instrument in lightly touching a finger at the midpoint or quarter point at either end of the string, for example,  in order to make it sound one or two  octaves higher than the fundamental, respectively.

 Finally, we note that this is the first treatment of the dramatic effect of superconducting substrates upon the output power of radiating BSCCO mesas.  Using the same Love principle that was the basis for the correct implementation of Amp{\`e}re's law, the magnetic equivalence principle, we conclude that the superconducting substrates cause a drastic reduction in the output power of the radiation, especially for the power emitted near to $\theta=90^{\circ}$.  Note that in our predictions for both the primary $ac$ Josephson current and the secondary cavity radiation sources, the output power at the fundamental frequency is predicted to be at least comparable to that at its maximum output.  The fact that in both rectangular and cylindrical mesas, the experimental output at $\theta=90^{\circ}$ is consistent with zero is a very strong indication that our magnetic conductor model of superconducting substrates is correct.  Our model  is also consistent with the experiments of Barbara {\it et al.}, in which the output of a Josephson junction array with the currents parallel to the superconducting substrate was enhanced by the substrate (or ground plane) prior to its entry into the waveguide\cite{Barbara1,Barbara2,Vasilic}.  We reiterate that removal of the superconducting substrate could enhance the output by at least two orders of magnitude, and by replacing the substrate with a perfect electric conductor such as Cu, one could further enhance the output by a factor of four.  This could allow for output as high as 5mW, which would be more than sufficient for many practical applications.

In summary, we have identified the primary microscopic source of the coherent radiation as the $ac$ Josephson current which locks onto a cavity mode, exciting it, and the two modes radiate together at the fundamental $ac$ Josephson frequency.  It suffices to treat all of the junctions as acting in unison.   For radiation at the fundamental $ac$ Josephson frequency locked onto the cylindrical cavity $(11)$ mode, we predict a combined output radiation pattern with linear electric field polarization that is usually along $\hat{\bm \theta}$, but not always.  The second harmonic in cylindrical mesas arises almost exclusively from the $ac$ Josephson current, and is also linearly polarized. the output from rectangular mesas should also be linearly polarized.  We reiterate that removal of the superconducting substrate, or better yet, replacement of it by a perfect electric conductor such as Au or Cu, could lead to an enhancement of the output power of the mesas from the highest observed power of 5$\mu$ W up to 5 mW, suitable for many applications.  We therefore name this device a Josephson STAR-emitter, for stimulated terahertz amplified radiation emitter.
\section{Acknowledgments}
We thank  X. Hu,  S. Lin, B. Markovic,  N. F. Pedersen, and M. Tachiki for stimulating discussions.  This work was supported in part both by the JST (Japan Science and Technology Agency) CREST project, by the WPI Center for Materials Nanoarchitechtonics (MANA),  by the JSPS (Japan Society for the Promotion of Science) CTC program and by the Grant-in Aid for Scientific Research (A) under the Ministry of Education, Culture, Sports, Science and Technology (MEXT) of Japan. One of us (R.A.K.) would also like to thank the University of Tsukuba for its kind hospitality.

\section{Appendix}

\subsection{A. Electric field from $\delta J_n({\bm x}')$}
We now calculate the electric field ${\bm E}_{{\bm A}_{\delta J}}({\bm x},t)$ arising from the $\delta J_n({\bm x})$ contribution to the electric surface current density. From Eq. (\ref{An}), we write,
\begin{eqnarray}
{\bm A}_{\delta J}({\bm x},t)&=&\frac{a\mu_0}{8\pi}\sum_{n=1}^{\infty}\int d^3{\bm x}'\hat{\bm z}'\eta(z')\delta(\rho'-a)\frac{e^{in(k_JR-\omega_Jt)}}{R}\nonumber\\
& &\times \sum_{p=1,m=0}^{\infty}C_{mp}^{(n)}J_m(k'_{mp}\rho')\cos[m(\phi'-\phi_{0n})].\nonumber\\
\end{eqnarray}
In the radiation zone, we have
\begin{eqnarray}
{\bm A}_{\delta J}({\bm x},t)&{{\rightarrow}\atop{r/a\gg1}}&-\hat{\bm\theta}\sin\theta\mu_0 v\sum_{n,p=1;m=0}^{\infty}C_{mp}^{(n)}(-i)^m\nonumber\\
& &\times\frac{e^{in(k_Jr-\omega_Jt)}}{4\pi r}J_m(k'_{mp}a)J_m(nk_{\theta})\nonumber\\
& &\times \cos[m(\phi-\phi_{0n})]S_n^J(\theta).\label{AdeltaJ}
\end{eqnarray}
We then find
\begin{eqnarray}
{\bm E}_{{\bm A}_{\delta J}}({\bm x},t)&{{\rightarrow}\atop{r/a\gg1}}&-i\hat{\bm\theta}\sin\theta\mu_0 v\sum_{n,p=1;m=0}^{\infty}C_{mp}^{(n)}n\omega_J(-i)^m\nonumber\\
& &\times\frac{e^{in(k_Jr-\omega_Jt)}}{4\pi r}J_m(k'_{mp}a)J_m(nk_{\theta})\nonumber\\
& &\times \cos[m(\phi-\phi_{0n})]S_n^J(\theta).\label{EdeltaJ}
\end{eqnarray}
Thus, combining this with ${\bm E}_{{\bm A}_J}({\bm x},t)$, we have
\begin{eqnarray}
{\bm E}_{{\bm A}}({\bm x},t)&{{\rightarrow}\atop{r/a\gg1}}&\frac{\hat{\bm\theta}v\mu_0\sin\theta}{4\pi r}\sum_{n=1}^{\infty}e^{in(k_Jr-\omega_Jt)}\nonumber\\
& &\times n\omega_JS^J_n(\theta)\sum_{m=0}^{\infty}(-i)^{m+1}E_{mn}J_m(nk_{\theta})\nonumber\\
& &\times \cos[m(\phi-\phi_{0n})],\label{EAfull}\\
E_{mn}&=&J_n^J\delta_{m,0}+\sum_{p=1}^{\infty}C_{mp}^{(n)}J_m(k'_{mp}a).
\end{eqnarray}
The $E_{mn}$ are real.
Thus, there are two effects, both of them we assume to be small, of including the full anisotropy of the spatial dependence of the $ac$ Josephson current into the surface electric current density source ${\bm J}_S$.  The first is a renormalization of the uniform $J_n^J$ coefficients to
\begin{eqnarray}
\tilde{J}_n^{J}&=&J_n^J+\sum_{p=1}^{\infty}C_{0p}^{(n)}J_0(k_{0p}'a),\label{renorm}
\end{eqnarray}
and the second is to introduce azimuthal anisotropy and additional axial anisotropy in the additional terms $J_m(nk_{\theta})\cos[m(\phi-\phi_{0n})]$.  We note that the renormalization in Eq. (\ref{renorm}) is a bit spurious, as the term would vanish if the integral were over the full volume of the cylinder, as noted following Eq. (\ref{deltaJform}),at least if the phase factor $e^{-i{\bm k}\cdot{\bm x}'}$ were absent, but we have placed the current only on the edge, modifying the integral.  The second change is more interesting.  It means that output power which is azimuthally anisotropic can arise from inhomogeneities in the $ac$ Josephson current itself, and does not necessarily depend upon the presence of the cavity.  However, we generally assume the $C_{mp}^{(n)}\ll J_n^J$, so that the primary effect of these spatial inhomogeneities $\delta J_n^J({\bm x}')$ is to excite a particular cavity mode, as discussed in Sec. IV.  Nevertheless, as discussed in Sec. VI, the case of $n=2$ is of particular interest, as Eq. (\ref{EAfull}) demonstrates that a $\phi$-dependence to the output power can arise from the inhomogeneous $ac$ Josephson current, although for every $n$, the output power from such terms is proportional to $\sin^2\theta$, vanishing at $\theta=0^{\circ}$.  As we shall see, if the $n=1$ $ac$ Josephson mode locks onto the $(11)$ cavity mode, the nearest cavity mode to the $n=2$ $ac$ Josephson harmonic is the $(01)$ mode, which should give rise to a very weak output, at best, but one that would be $\phi$-independent, as well.  Another point of some interest is that the $m$th contribution to ${\bm E}_{\bm A}$ is proportional to $(-1)^{m+1}$.  This could be important when the radiation originates both from the inhomogeneous part of the $ac$ Josephson ${\bm J}_S$ and cavity ${\bm M}_S$ sources, as discussed in Sec. VI.
\subsection{B. Resonant and off-resonant cavity radiation}
The ${\bm F}$ vector potential resulting from the resonant cavity mode is then
  \begin{eqnarray}
  {\bm F}({\bm x},t)&{\rightarrow\atop r/a\gg1}&\frac{iv\epsilon_0}{4\pi r}\omega_{mp}A^{(n)}_{mp}(\infty)S_{mp}^M(\theta)e^{i(k_{mp}r-\omega_{mp}t)}\nonumber\\
  & &\times J_{m}(k'_{mp}a)\int_0^{2\pi}\frac{d\phi'}{2\pi}\hat{\bm\phi}'\cos[m(\phi'-\phi_{0n})]\nonumber\\
  & &\times e^{-ik_{mp}a\sin\theta\cos(\phi-\phi')}\Bigr|_{{p=p_0}\atop{m=m_0}},
  \end{eqnarray}
  where $\omega_{m_0p_0}=n\omega_J$ and $k_{m_0p_0}=\omega_{m_0p_0}/c$ outside the mesa, and $S_{mp}^M(\theta)=1$ when the mesa is suspended in vacuum.
  After setting $\hat{\bm{\phi}}'=\hat{\bm{\theta}}\cos\theta\sin(\phi-\phi')+\hat{\bm{\phi}}\cos(\phi-\phi')$, and evaluating the integral over $\phi'$, we obtain
  \begin{eqnarray}
  {\bm F}({\bm x},t)&{\rightarrow\atop r/a\gg1}&-(-i)^{m-1}\frac{G_{mp}^{(n)}\epsilon_0}{k_{mp}}\frac{e^{i(k_{mp}r-\omega_{mp}t)}}{4\pi r}\nonumber\\
  & &\times\Bigl[\hat{\bm\theta}\cos\theta\sin[m(\phi-\phi_{0m})]J_m^{+}(k_{mp}^\theta)\nonumber\\
  & &\>\>-\hat{\bm\phi}\cos[m(\phi-\phi_{0n})]J_m^{-}(k_{mp}^{\theta})\Bigr]\biggr|_{{p=p_0}\atop{m=m_0}},\label{Fresonant}\nonumber\\
  \end{eqnarray}
  where
  \begin{eqnarray}
  G^{(n)}_{mp}&=&\frac{v}{2\epsilon}k_{mp}t_{\rm eff}C_{mp}^{(n)}J_m(k'_{mp}a)S_{mp}^M(\theta),\label{Gmp}
  \end{eqnarray}
  $k_{mp}^{\theta}=k_{mp}a\sin\theta$, and $J_{m}^{\pm}(z)=[J_{m+1}(z)\pm J_{m-1}(z)]/2$.
  The same procedure can be followed for the ${\bm F}$ vector potential arising from the off-resonant $\omega=n\omega_J$ harmonic electric current source exciting the tail of an $(mp)$ cavity mode, where $(mp)\ne(m_0p_0)$, and Eqs. (\ref{Fresonant}) and (\ref{Gmp}) are again obtained, with the replacements that $\omega_{mp}\rightarrow n\omega_J$, $k_{mp}\rightarrow nk_J=n\omega_J/c$ in Eq. (\ref{Fresonant}), and $G_{mp}^{(n)}\rightarrow -i\tilde{G}^{(n)}_{mp}$, where
  \begin{eqnarray}
  \tilde{G}^{(n)}_{mp}&=&\frac{vn^2k_J\omega_JC_{mp}^{(n)}}{\epsilon[\omega_{mp}^2-(n\omega_J)^2]}J_m(k'_{mp}a)S_n^M(\theta),\label{tildeGmp}
  \end{eqnarray}
  where the substrate factor $S_n^M(\theta)=1$ when the mesa is suspended in vacuum, and otherwise differs only from $S_{mp}^M(\theta)$ in that $k_{mp}\rightarrow nk_J$, as noted in Sec. VIII.
  \subsection{C. Vector potentials and surface currents for rectangular mesas}
  For the rectangular mesas studied in Sec. IX,
   ${\bm A}({\bm x},t)$ and ${\bm F}({\bm x},t)$ in spherical coordinates are then given in the radiation zone by
\begin{eqnarray}
{\bm A}({\bm x},t)&\rightarrow&\frac{\mu_0\hat{\bm z}J_J\tilde{v}}{8\pi r}\sum_{n=1}^{\infty}a_ne^{in(k_Jr-\omega_Jt)}S^J_n(\theta)\chi_n,\label{Anrect}\\
\chi_n(\theta,\phi)&=&\cos X_n\frac{\sin Y_n}{Y_n}+\cos Y_n\frac{\sin X_n}{X_n},\label{chin}\\
{\bm F}({\bm x},t)&\rightarrow&-\frac{\epsilon_0\tilde{v}}{16\pi r}\sum_{n=1}^{\infty}\tilde{E}_{0n}e^{in(k_Jr-\omega_Jt)}S^M_n(\theta)\nonumber\\
& &\times
(\hat{\bm x}M^x_{n}+\hat{\bm y}M^y_{n}),\\
M_n^x&=&-\sin Y_n\sum_{\sigma=\pm}\sigma e^{i\sigma n\pi x_n/w}\frac{\sin(X_{n,\sigma})}{X_{n,\sigma}},\\
M_n^y&=&\frac{\sin Y_n}{Y_n}\sum_{\sigma=\pm}e^{i\sigma X_n}\sin\Bigl(\frac{n\pi}{2}+\frac{\sigma n\pi x_n}{w}\Bigr),\label{Mny}
\end{eqnarray}
where $X_n=(k_nw/2)\sin\theta\cos\phi$,
$Y_n=(k_n{\ell}/2)\sin\theta\sin\phi$, $k_nw=n\pi/n_r$, $X_{n,\sigma}=n\pi/2+\sigma X_n$, $\tilde{v}=w{\ell}h$, $S^M_n(\theta)=S^J_n(\theta)=1$ for no substrate,  $\hat{\bm x}=\hat{\bm r}\sin\theta\cos\phi+\hat{\bm\theta}\cos\theta\cos\phi-\hat{\bm\phi}\sin\phi$ and $\hat{\bm y}=\hat{\bm r}\sin\theta\sin\phi+\hat{\bm\theta}\cos\theta\sin\phi+\hat{\bm\phi}\cos\phi$, and $x_n$ appears in Eq. (\ref{Mrect}).

The quantities present in the expression for the  output power intensity from the combined electric and magnetic surface current density sources follow.  For the portion of the output power resulting from the uniform portion of ${\bm J}_S$, Eqs. (\ref{Anrect}) and (\ref{chin}) are sufficient.
The part of the combined output arising from ${\bm M}_S$ is more complicated, and the combination can be either coherent or incoherent, as for the output from cylindrical mesas. When the combination is coherent, we assume that $H_y(x'=\pm w/2)=0$ is maintained along both lengths of the mesa, but the output power $P(x_n)$  is evaluated when $x_n$ is either $0$ or $w/n$ for $n$ odd, or when $x_n=\pm w/2n$ for $n$  even.  When the output from the two sources is incoherent with respect to one another, then we average $P(x_n)$ in two models.  In Model I, we also assume that $H_y(x'=\pm w/2)=0$ is maintained along both lengths of the mesa, so that $\langle P(x_n)\rangle_{I}=\frac{1}{2}[P(0)+P(w/n)]$ for $n$ odd, and $\langle P(x_n)\rangle_{I}=\frac{1}{2}[P(w/2n)+P(-w/2n)]$ for $n$ even.  In Model II, we relax the boundary condition upon $H_y$, and average over all $x_n$ values that preserve the wave vector within the mesa, $\langle P(x_n)\rangle_{II}=(n/2w)\int_{-w/n}^{w/n}P(x_n)dx_n$.  In both models, the average output power is characterized by three functions $C_n^i(\theta,\phi)$,  $D_n^i(\theta,\phi)$, and $E_n^i(\theta,\phi)$ for $i=$ I and II.
For Model I, $C_n^{I}=A_n^2$, $D_n^I=B_n^2$, and $E_n^I=2A_nB_n$, where for $n$ either odd or even,
 \begin{eqnarray}
 A_n&=&\sin Y_n\sum_{\sigma=\pm}\frac{\sigma^n\sin(X_{n,\sigma})}{X_{n,\sigma}},\\
 B_n&=&2\frac{\sin Y_n}{Y_n}\sin\Bigl(\frac{n\pi}{2}+X_n\Bigr).
 \end{eqnarray}
 For Model II,
 \begin{eqnarray}
 C_n^{II}&=&\sin^2Y_n\sum_{\sigma=\pm}\Bigl(\frac{\sin(X_{n,\sigma})}{X_{n,\sigma}}\Bigr)^2,\\
 D_n^{II}&=&\frac{\sin^2Y_n}{Y^2_n}\Bigl(1-(-1)^n\cos(2X_n)\Bigr),\\
 E_n^{II}&=&2\frac{\sin^2Y_n}{Y_n}\sum_{\sigma=\pm}\frac{\sigma\sin^2(X_{n,\sigma})}{X_{n,\sigma}},
 \end{eqnarray}
 where $X_n$, $Y_n$, and $X_{n,\sigma}$ are given following Eq. (\ref{Mny}).


\begin{thebibliography}{99}
\bibitem{note} Hu and Lin in their Fig. 6 also found results of the radiation patterns from the cylindrical cavity $(11), (12)$, $(01)$, and $(21)$ modes.  Although their paper described numerical calculations inside the mesa, their results essentially agree with ours for the radiation-zone output power emanating from the cavity modes alone from a sample suspended in vacuum. They made no mention of their techniques in obtaining these figures, however, whereas we gave analytic expresssions for them.  They also used the same boundary condition as we did for the cavity excitations, ${H}_{\phi}=0$, but they did not include the electric surface current ${\bm J}_S$ which necessarily occurs from the combination of the presence of the $ac$ Josephson current and the boundary assumption $H_{\phi}=0$\cite{antenna,antenna2}, violating Love's magnetic equivalence principle.  They also made no mention of the superconducting substrate, and did not predict any higher harmonics to be observable, so their results are in qualitative disagreement with experiment\cite{Kadowaki3}.
\end{thebibliography}
\end{document}